\documentclass[12pt]{article}

\usepackage{amssymb,amsmath,amsthm}
\usepackage{array,longtable}
\usepackage[dvips]{graphicx}
\usepackage[dvips]{color}
\usepackage[mathscr]{eucal}

\newcolumntype{V}{>{$}m{4cm}<{$}}
\newcolumntype{C}{>{$}c<{$}}
\newcolumntype{L}{>{$}l<{$}}
\newcolumntype{R}{>{$}r<{$}}

\newcommand{\Rh}{\mathbb R}
\newcommand{\Zh}{\mathbb Z}

\newcommand{\Nc}{\mathcal{N}}

\newcommand{\pd}{\partial}

\newcommand{\arctanh}{\mathop{\mathrm{arctanh}}\nolimits}

\newcommand{\ap}{\alpha^{\prime}}

\allowdisplaybreaks[3]

\renewcommand{\theequation}{\arabic{section}.\arabic{equation}}

\begin{document}
\title{
$~$\\
\textsc{Diagonal Representation of Open String Star}
\\
\textsc{and}
\\
\textsc{Moyal Product}
$~$\\
$~$\\}
\author{
\textsf{D.M.~Belov}\footnote{On leave from Steklov Mathematical
Institute, Moscow, Russia.}
\vspace{3mm}
\\
Department of Physics
\\
Rutgers University
\\
136 Frelinghuysen Rd.
\\
Piscataway, NJ 08854, USA
\vspace{3mm}
\\
\texttt{belov@physics.rutgers.edu}
}

\date{~}
\maketitle
\thispagestyle{empty}

\begin{abstract}
We explicitly find the spectrum of the operators $M^{rs}$
and $\widetilde{M}^{rs}$, which specify the $\star$-product
in the matter and ghost sectors correspondingly.
Further we derive the diagonal representation for the
$3$-string vertices in both sectors. Using this representation
we identify the appearing Moyal structures in the matter sector.
In addition to the continuous non-commutativity parameter
$\theta(\kappa)$ found in \cite{0202087} we find
the discrete non-commutativity parametrized by $\theta_{\xi}$.
\end{abstract}

\newpage

\tableofcontents

\section{Introduction}\label{sec:intro}
\setcounter{equation}{0}
Recently it has been a resurgence of interest in string
field theory (SFT) \cite{OSFT}. Along with the applications to such problems
as the tachyon condensation \cite{sen}, \cite{0102085,0109182,ABGKM} the
intrinsic structure of SFT is also investigated.
In particular recently
M.~Douglas, H.~Liu, G.~Moore and B.~Zwiebach
\cite{0202087}
have shown
that open zero-momentum string field algebra
 can be represented as a continuous product of mutually commuting
Moyal algebras. The first attempt to describe the open string
star product as a series of Moyal products was made in \cite{0106157}.
The aim of the present paper is to generalize these results
to the matter sector including zero modes and to the ghost sector.
One hopes that finding such a diagonal representation will lead
to a better understanding of string field $\star$-algebra structure and
finding the solutions of SFT equations of motion.

\vspace{1cm}

The paper is organized as follows.
In Section~\ref{sec:3string} we review the structure of $3$-string
vertex in the matter and ghost sectors.
In Section~\ref{sec:spectrum} we find the spectrum
of operators $CU$ and $UC$ defining the Neumann matrices in the matter sector.
Subsection~\ref{sec:summary}
contains summary on the spectrum of the matter operator $CU$.
In Section~\ref{sec:spectrum-gh} we consider the
spectrum of various operators defining
the ghost Neumann matrices.
In Section~\ref{sec:diag} we derive the diagonal representation
for the matter and
ghost $3$-string vertices.
In Section~\ref{sec:moyal} identify the Moyal structures
appearing in the diagonal representation of the vertices.

Appendices contain necessary technical information.
In Appendix~\ref{app:conventions} we present
the normalization conventions for the coordinate
and momentum eigenstates.
In Appendix~\ref{app:spectrum} we present a detailed
derivation of the spectrum of matter operator $CU$.

\section{Review of the $3$-string vertex}
\label{sec:3string}
\setcounter{equation}{0}



Let us remind the expressions for the $3$-string vertices in
the matter and ghost sectors \cite{GJ1,GJ2} in the presence of the
constant background metric $g_{\mu\nu}$:
\begin{subequations}
\begin{align}
|V_3^{\text{m}}\rangle_{123}&=\Nc_m^{-26}\exp\Bigl[-\frac12\sum_{r,s=1}^{3}\sum_{m,n=0}^{\infty}
(CM^{rs})_{nm} (a_{n}^{(r)\dag},\,a_{m}^{(s)\dag})
\Bigr]|\Omega_b\rangle_{123},
\label{V3:matter}
\\
|V_3^{\text{gh}}\rangle_{123}&=\exp\Bigl[-\sum_{r,s=1}^{3}\sum_{n=0,m=1}^{\infty}
b_{-n}^{(r)}\Bigl(\frac{1}{\sqrt{N}}(C\tilde{M}^{rs})\sqrt{N}\Bigr)_{nm} c_{-m}^{(s)}
\Bigr]|+\rangle_{123},
\label{V3:ghost}
\end{align}
\label{V3}
\end{subequations}
where
\begin{subequations}
\begin{align}
CM^{rs}&=\frac{1}{3}(C+\alpha^{s-r} U+\alpha^{r-s}\overline{U}),
\qquad U^{\dag}=U,\quad \overline{U}=CUC,\quad U^2=1;
\\
C\tilde{M}^{rs}&=\frac{1}{3}(C+\alpha^{s-r} \widetilde{U}+\alpha^{r-s}\overline{\widetilde{U}}),
\qquad \widetilde{U}^{\dag}=\widetilde{U},\quad
\overline{\widetilde{U}}=C\widetilde{U}C,\quad \widetilde{U}^2=1
\label{tUrs}
\end{align}
and matrices $N$ and $C$ are given by
\begin{equation}C_{mn}=(-1)^m\delta_{mn}\quad\text{and}\quad
N_{nm}=n\delta_{n,m}+\delta_{m,0}\delta_{n,0}.
\end{equation}\end{subequations}
By the bar over the matrix $\overline{U}$ we assume the complex conjugation.
The vacuum states are defined as follows
\begin{subequations}
\begin{align}
a_n&|\Omega_b\rangle=0,\quad n\geqslant 0;
\\
b_n&|+\rangle=c_n|+\rangle=0,\quad n\geqslant 1\quad\text{and}\quad c_0|+\rangle=0.
\end{align}
\end{subequations}
The zero mode oscillators $a_{0,\mu}$ and $a_{0,\mu}^{\dag}$ are related to the
coordinate and momentum modes in the open string expansion as
\begin{equation*}x_{\mu}=i\frac{\sqrt{\ap b}}{2}(a_{0,\mu}-a_{0,\mu}^{\dag})
\quad\text{and}\quad
p_{\mu}=\frac{1}{\sqrt{b\ap}}(a_{0,\mu}+a_{0,\mu}^{\dag}).
\end{equation*}Here $b$ is a real parameter introduced in \cite{0102112}.

In general the normalization factor $\Nc_m$ appearing in formula \eqref{V3:matter}
depends on the normalization of the coordinate and momentum
eigenstates $|x\rangle$ and $|p\rangle$. We will use the normalization of these
sates described in Appendix~\ref{app:conventions}, for which $\Nc_m$ is of the form
\begin{equation}\Nc_m=\frac{\sqrt{3}}{b}\left(V_{00}+\frac{b}{2}\right)(2\pi\ap b)^{\frac{1}{4}}.
\end{equation}

The operators $CM$ and $C\tilde{M}$ are not independent,
there is a relation between them. This relation
is very much based on the
associativity of the Witten's $\star$-product, and
can be obtained from the consideration of $4$-string vertices in the matter
and ghost sectors. The $4$-string
vertices in matter and ghost sectors are defined by the matrices $V$ and $\widetilde{V}$
correspondingly. As shown in \cite{GJ2} these matrices are related to each other by
$\widetilde{V}=-\overline{V}$.

The relation between the matrices $V$ and $U$,
which define $4$-string and $3$-string vertices correspondingly,
is given by \cite{GJ1}
\begin{subequations}
\begin{align}
V+\overline{V}&=4(1+C(U+\overline{U}))[U+\overline{U}+4C]^{-1},
\\
V-\overline{V}&=2\sqrt{3}(U-\overline{U})C[U+\overline{U}+4C]^{-1}.
\end{align}
\label{VU-rel}
\end{subequations}
The solution of these equations is the following
\begin{subequations}
\begin{align}
U+\overline{U}&=-4(1-C(V+\overline{V}))[4C-V-\overline{V}]^{-1},
\\
U-\overline{U}&=-2\sqrt{3}C(V-\overline{V})[4C-V-\overline{V}]^{-1}.
\end{align}
\label{UV-rel}
\end{subequations}
These expressions define the function $U=U(V,\overline{V})$.
Similar relations to \eqref{UV-rel} can be obtained for the
matrices $\widetilde{U}$ and $\overline{\widetilde{U}}$, one only
needs to change $V\mapsto\widetilde{V}=-\overline{V}$ \cite{GJ1}. And therefore
$\widetilde{U}=U(-\overline{V},-V)$\footnote{Notice that
these relation is only true for $b=2$.} or more precisely
\begin{subequations}
\begin{align}
\widetilde{U}+\overline{\widetilde{U}}&=-4(1+C(V+\overline{V}))[4C+V+\overline{V}]^{-1},
\\
\widetilde{U}-\overline{\widetilde{U}}&=
2\sqrt{3}C(V-\overline{V})[4C+V+\overline{V}]^{-1}.
\end{align}
\label{tildeUV-rel}
\end{subequations}

For the spectroscopy problem it is more convenient to consider
matrices $CU$ and $UC$ instead of matrices $U$ and $\overline{U}$.
The reason to do this is that operator $CU$ is a unitary operator.
The equations \eqref{UV-rel} and \eqref{tildeUV-rel} allow us
to express the operators $C\tilde{U}$ and $\tilde{U}C$ in terms
of the operators $CU$ and $UC$. This relation is of the form
\begin{equation}
C\tilde{U}=-\frac{2CU+1}{CU+2}\quad\text{and}\quad
\tilde{U}C=-\frac{2UC+1}{UC+2}.
\end{equation}
From these relations it follows that eigenvalues
of the operators $C\tilde{U}$ and $CU$ are related by
the $PSL(2,\Zh)$ transformation $P$
\begin{equation}P(z)=-\frac{2z+1}{z+2}=-2+\frac{3}{z+2}.
\label{P(z)}
\end{equation}
The transformation $P$ satisfy the equation $(P\circ P)(z) =z$.

\section{Spectrum of the operators $CU$ and $UC$ (matter)}
\label{sec:spectrum}
\setcounter{equation}{0}



\subsection{Notations}
\label{sec:spectrum:not}
We have the followings matrices $C'_{mn},\,C_{ab},\,U'_{mn}$ and $U_{ab}$,
where $a,b\geqslant 0$ and $m,n\geqslant 1$. The operators $C$ and $C'$
define BPZ conjugation and are given by $C_{ab}=(-1)^a\delta_{ab}$
and $C'_{mn}=(-1)^m\delta_{mn}$.
There operators $U$ and $U'$ have the following properties:
\begin{equation}
U^{\prime\dag}=U',\quad \overline{U'}=C'U'C',\quad U^{\prime 2}=1
\qquad\text{and}\qquad
U^{\dag}=U,\quad \overline{U}=CUC,\quad U^2=1,
\end{equation}
where by $\overline{U}$ we denote complex conjugated operator
to $U$.
The operators $U$ and $U'$ are related by
\begin{equation}U'_{mn}=U_{mn}+\frac{1}{1-U_{00}}U_{m0}U_{0n}
\quad\text{and}\quad U_{00}=1-\frac{b}{V_{00}+\frac{b}{2}}.
\label{relUU'1}
\end{equation}The vector $U_{m0}$ is an eigenvector of the matrix $U'$ with
eigenvalue $+1$:
\begin{equation}\sum_{n=1}^{\infty}U'_{mn}U_{n0}=U_{m0},
\qquad
\sum_{m=1}^{\infty}U_{0m}U_{m0}=1-U^{2}_{00}.
\label{Uon}
\end{equation}The vectors $U_{m0}$ and $U_{0m}$ are $b$-dependent,
therefore it is convenient to introduce
$b$-independent vectors $W_n$ and $\overline{W}_n$ \cite{0102112} as
\begin{equation}W_n=\frac{\sqrt{b}}{1-U_{00}}\, U_{0n}\quad\text{and}\quad
\overline{W}_n\equiv (-1)^nW_n=\frac{\sqrt{b}}{1-U_{00}}\, U_{n0}.
\label{wWWw}
\end{equation}We also introduce the scalar product (notice that our definition involves
complex conjugation):
\begin{equation}(W,W)\equiv \sum_{n=1}^{\infty}\overline{W}_nW_n
\stackrel{\ref{Uon}}{=}2V_{00}.
\end{equation}The expression \eqref{relUU'1} takes the following
form in terms of $W$'s:
\begin{equation}U'_{mn}=U_{mn}+\frac{\overline{W}_mW_n}{V_{00}+\frac{b}{2}}.
\label{relUU'}
\end{equation}

\begin{figure}
\centering
\includegraphics[width=120pt]{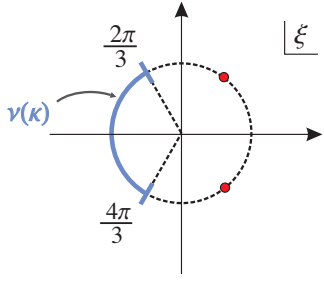}
\caption{On the picture we have presented the spectrum of
the operator $CU$. It consist of continuous spectrum,
which is represented by the function $\nu(\kappa)$ (solid blue line),
and discrete spectrum, which contains two points $\xi$ and $\overline{\xi}$
(red dots).}
\label{fig:1}
\end{figure}

The matrices $(C'U')_{mn}$ and $(U'C')_{mn}$ have common
eigenvectors $v^{(\kappa)}_n$:
\begin{equation}\sum_{n=1}^{\infty}(C'U')_{mn}v^{(\kappa)}_{n}=\nu(\kappa)v^{(\kappa)}_{m}
\quad\text{and}\quad
\sum_{n=1}^{\infty}(U'C')_{mn}v^{(\kappa)}_{n}=\overline{\nu}(\kappa)v^{(\kappa)}_{m},
\end{equation}where the components of the vectors $v^{(\kappa)}_m$ are explicitly
given by the following generating function \cite{0111281}
\begin{equation}f^{(\kappa)}(z)=\sum_{m=1}^{\infty}v_m^{(\kappa)}\frac{z^m}{\sqrt{m}}
=\frac{1}{\kappa}(1-e^{-\kappa\arctan z}).
\label{gen-func}
\end{equation}Note that since $(C'U')(U'C')=1$ we get that $|\nu(\kappa)|^2=1$.
To get an explicit expression for the eigenvalues $\nu(\kappa)$
we will use the eigenvalues $\mu^{rs}(\kappa)$ of
the operators $M^{\prime rs}$ \cite{0111281}
\begin{subequations}
\begin{align}
\mu^{rs}(\kappa)&=\frac{1}{1+2\cosh\frac{\pi\kappa}{2}}
\Bigl[1-2\delta_{r,s}+e^{\frac{\pi\kappa}{2}}\delta_{r+1,s}
+e^{-\frac{\pi\kappa}{2}}\delta_{r,s+1}\Bigr],
\\
M^{\prime rs}&=\frac{1}{3}\Bigl[1+\alpha^{s-r}C'U'+\alpha^{r-s}U'C'\Bigr].
\label{exM'}
\end{align}
\label{nuonu}
\end{subequations}
From the formulae \eqref{nuonu} one can obtain the following
expression for the eigenvalue $\nu(\kappa)$
\begin{equation}\nu(\kappa)=-\frac{1}{1+2\cosh\frac{\pi\kappa}{2}}\Bigl[
2+\cosh\frac{\pi\kappa}{2}+i\sqrt{3}\sinh\frac{\pi\kappa}{2}
\Bigr].
\label{nu(kappa)}
\end{equation}The analysis of this formula shows that
 $\frac{2\pi}{3}< \arg \nu(\kappa)< \frac{4\pi}{3}$
(see Figure~\ref{fig:1}).

The eigenvectors $v^{(\kappa)}$ belong to the continuous spectrum.
As it follows from the expression \eqref{nu(kappa)} the spectrum
of operators $C'U'$ and $U'C'$ is non-degenerate.

The normalization of the vectors $v^{(\kappa)}$ is given by \cite{0201015}
\begin{equation}(v^{(\kappa)},\,v^{(\kappa')})
=\Nc(\kappa)\delta(\kappa-\kappa'),
\quad\text{where}\quad \Nc(\kappa)=\frac{2}{\kappa}\sinh\frac{\pi\kappa}{2}.
\label{N(kappa)}
\end{equation}
Since the operator $M^{\prime 11}$ \eqref{exM'} is a hermitian operator and
$v^{(\kappa)}$ are its only eigenvectors \cite{0111281} we conclude that
the vectors $v^{(\kappa)}$ form a complete set:
\begin{equation}\delta_{mn}=\int d\kappa \frac{1}{\Nc(\kappa)}\, v^{(\kappa)}_m v^{(\kappa)}_n.
\label{unity}
\end{equation}
The scalar products among $W_n$ and $v_n^{(\kappa)}$ can be obtained
from the paper\footnote{To this end one have to use
expression (3.11) from \cite{0201136} and the following relation
between $v_e,\,v_o$ and $W$:
$$
W=-\sqrt{2}(v_e-iv_o)\quad\text{and}\quad
\overline{W}=-\sqrt{2}(v_e+iv_o).
$$
Notice that this relation is different from the one presented in \cite{0201136}.
The origin of these difference is explained in the end of Appendix~A of \cite{0102112}.
} \cite{0201136} and are given by
\begin{subequations}
\begin{align}
(v^{(\kappa)},\,W)&=-\frac{\sqrt{2}}{\kappa}\,
\frac{\cosh\frac{\pi\kappa}{2}-1-i\sqrt{3}\sinh\frac{\pi\kappa}{2}}{2\cosh\frac{\pi\kappa}{2}+1}
\equiv-\frac{\sqrt{2}}{\kappa}[1+\nu(\kappa)].
\\
(v^{(\kappa)},\,\overline{W})&=
-\frac{\sqrt{2}}{\kappa}\,
\frac{\cosh\frac{\pi\kappa}{2}-1+i\sqrt{3}\sinh\frac{\pi\kappa}{2}}{2\cosh\frac{\pi\kappa}{2}+1}
\equiv-\frac{\sqrt{2}}{\kappa}[1+\overline{\nu}(\kappa)],
\label{vWscalar}
\end{align}
\end{subequations}

\subsection{Eigenvectors of the operators $CU$ and $UC$.}
\label{sec:spectrum:eigen}
We want to find a spectrum of the operators $CU$ and $UC$.
Since $UC=(CU)^{\dag}$ it is enough to find the spectrum of
the operator $CU$. To this end we have to solve the equation
\begin{equation}\sum_{b=0}^{\infty}(CU)_{ab}V_b^{(\xi)}=\xi V_{a}^{(\xi)}.
\end{equation}Notice that since $CU$ is a unitary operator all of its eigenvalues
$\xi$ belong to the unit circle.
Consideration of the two cases $a\ne 0$ and $a=0$ yields the
equations:
\begin{subequations}
\begin{align}
& \sum_{n=1}^{\infty} (CU)_{mn}V_{n}^{(\xi)}
+\sum_{n=1}^{\infty}C_{mn}U_{n0}V_0^{(\xi)}=\xi V_m^{(\xi)},
\label{eqA}
\\
& \sum_{n=1}^{\infty}U_{0n}V_n^{(\xi)}+U_{00}V_0^{(\xi)}=\xi V_0^{(\xi)}.
\label{eqB}
\end{align}
\end{subequations}
Using the expression \eqref{relUU'} one \eqref{wWWw} one can rewrite
the equation \eqref{eqA} in the following form:
\begin{equation*}\sum_{n=1}^{\infty} \Bigl[(C'U')_{mn}-\xi\delta_{mn}\Bigr]V_{n}^{(\xi)}
=\frac{1}{\sqrt{b}}W_m\,\sum_{n=1}^{\infty}U_{0n}V_n^{(\xi)}
-\frac{1-U_{00}}{\sqrt{b}}W_m V_0^{(\xi)}.
\end{equation*}Further simplification can be achieved by use of
equation \eqref{eqB}.
Finally the equations we have to solve get the following simple form
\begin{subequations}
\begin{align}
&\sum_{n=1}^{\infty} \Bigl[(C'U')_{mn}-\xi\delta_{mn}\Bigr]V_{n}^{(\xi)}
=W_m \frac{V_0^{(\xi)}}{\sqrt{b}}(\xi-1),
\label{EQa}
\\
& (\overline{W},V^{(\xi)})=\frac{V_{0}^{(\xi)}}{\sqrt{b}}\Bigl[
\Bigl(V_{00}+\frac{b}{2}\Bigr)\xi-\Bigl(V_{00}-\frac{b}{2}\Bigr)
\Bigr].
\label{EQb}
\end{align}
\label{EQs}
\end{subequations}

To solve these equations it is very important to notice that
the operator $C'U'$ has only continuous spectrum. In this case
one can easily write the solution to the equation \eqref{EQa}.
We have to distinguish two different cases:
\begin{enumerate}
\item The number $\xi$ is in the spectrum of $C'U'$. This means
that there is a real number $\kappa$ such that $\xi=\nu(\kappa)$.
In this case the solution of the \eqref{EQa} is given by
\begin{subequations}
\begin{equation}V_m^{(\kappa)}=A^{(\kappa)} v_m^{(\kappa)}-[1-\nu(\kappa)]
\frac{V_{0}^{(\kappa)}}{\sqrt{b}}
\,\left(\mathscr{P}\frac{1}{C'U'-\nu(\kappa)}\right)_{mn}W_n,
\label{cont:V}
\end{equation}where $A^{(\kappa)}$ is an arbitrary number and
$\mathscr{P}\frac{1}{x}$ means principal value.
The only reason why we have chosen the principal value
instead of for example $\frac{1}{x+i0}$ is
that
the formulaes will be a bit simpler for this case.
Substitution of this solution
into \eqref{EQb} allows us to determine $A^{(\kappa)}$ in terms
of $V_0^{(\kappa)}$:
\begin{equation}A^{(\kappa)}
=\frac{V_0^{(\kappa)}}{\sqrt{b}}
\frac{\nu(\kappa)-1}{(\overline{W},\,v^{(\kappa)})}
\left[V_{00}+\frac{b}{2}\frac{\nu(\kappa)+1}{\nu(\kappa)-1}
-(\overline{W},\,\mathscr{P}\frac{1}{C'U'-\nu(\kappa)}\,W)
\right]
\end{equation}\label{cont:A}
\end{subequations}
As we will see equations \eqref{cont:A} define a continuous
spectrum of the operator $CU$.
This continuous spectrum is
\textit{non-degenerate} and consists of the vectors $V^{(\kappa)}$ with
eigenvalues $\nu(\kappa)$.

\item If $\xi$ is not in the spectrum of $C'U'$, the solution
of the equation \eqref{EQa} is given by
\begin{subequations}
\begin{equation}V_m^{(\xi)}=\frac{V_{0}^{(\xi)}}{\sqrt{b}}
(\xi-1)\,\left(\frac{1}{C'U'-\xi}\right)_{mn}W_n.
\label{EQ-Vdisc}
\end{equation}Substitution of this solution
into \eqref{EQb} yields the equation on the eigenvalue $\xi$ (discrete spectrum):
\begin{equation}(\overline{W},\,\frac{1}{C'U'-\xi}\,W)=V_{00}+\frac{b}{2}\frac{\xi+1}{\xi-1}.
\label{EQ-desc}
\end{equation}\end{subequations}
This equation in general has two complex conjugated solutions belonging
to the unit circle (see Figure~\ref{fig:1}). One can easily check that
the vector $CV^{(\xi)}$ is also an eigenvector
of the operator $CU$ with the eigenvalue $\overline{\xi}$.
\end{enumerate}

The details of solving of these equations are presented
in Appendix~\ref{app:spectrum}. The results are collected
in the next subsection.

\subsection{Summary}
\label{sec:summary}
The spectrum of the operators $CU$ and $UC$ consist
of two branches: continuous and discrete.
\begin{itemize}
\item \textbf{Discrete spectrum}\\
The discrete spectrum of the operator $CU$ consist of
two points $\xi$ and $\overline{\xi}$ (see Figure~\ref{fig:1}). They are given
as the solutions of the equation
\begin{equation}\Re F(\eta)=\frac{b}{4},
\label{eta(b)}
\end{equation}where
\begin{equation*}
F(\eta)=\psi(\tfrac{1}{2}+\tfrac{\eta}{2\pi i})-\psi(\tfrac{1}{2})
\end{equation*}
and $\xi$ is related to $\eta$ by
\begin{equation}\xi=-\frac{1}{1-2\cosh\eta}\Bigl[
2-\cosh\eta-i\sqrt{3}\sinh\eta
\Bigr].
\label{sum:xi}
\end{equation}
For $b>0$ equation \eqref{eta(b)} has two real solutions.
On the Figure~\ref{fig:3}
we presented the $\arg\bar{\xi}(b)$ for this solution.

The generating functions for the eigenvectors from the discrete
spectrum is given by
\begin{subequations}
\begin{multline}
F^{(\xi)}(z)\equiv\sum_{n=1}^{\infty}V_n^{(\xi)}\frac{z^n}{\sqrt{n}}
=-\sqrt{\frac{2}{b}}\,V_0^{(\xi)}\left[
\frac{b}{4}+\frac{\pi}{2\sqrt{3}}\frac{\xi-1}{\xi+1}
+\log iz
\right.
\\
\left.
+e^{-2i(1+\frac{\eta}{\pi i})\arctan z}
\mathrm{LerchPhi}\bigl(e^{-4i\arctan z},1,\tfrac12+\tfrac{\eta}{2\pi i}\bigr)
\right],
\label{summary:gendisc}
\end{multline}
where
$$
x\mathrm{LerchPhi}(w,1,x)=\,{}_2F_1(1,x;x+1;w).
$$
The action of the operator $C$ on the eigenvectors
can be read from the formulae
\begin{equation}F^{(\overline{\xi})}(z)\equiv F^{(\xi)}(-z)
\quad\Longleftrightarrow\quad
CV^{(\xi)}=V^{(\overline{\xi})}.
\end{equation}The norm of the vectors from the discrete spectrum is given by
\begin{equation}(\!(V^{(\xi)},V^{(\xi)})\!)=|V_0^{(\xi)}|^2\sinh\eta \frac{\pd}{\pd\eta}
\Bigl[\log \Re F(\eta)\Bigr].
\label{norm:d}
\end{equation}\end{subequations}
Further we will assume that $V_0^{(\xi)}$ is chosen in such way (see \eqref{V0:disc})
that the norm \eqref{norm:d} is equal to one.

\item \textbf{Continuous spectrum}\\
The continuous spectrum of the operator $CU$ is given
by the function $\nu(\kappa)$, $\kappa\in\Rh$ (see Figure~\ref{fig:1})
\begin{equation}\nu(\kappa)=-\frac{1}{1+2\cosh\frac{\pi\kappa}{2}}
\left[2+\cosh\frac{\pi\kappa}{2}+i\sqrt{3}\sinh\frac{\pi\kappa}{2}\right].
\label{sum:nu}
\end{equation}The generating function for the eigenvector corresponding to eigenvalue
$\nu(\kappa)$ is given by
\begin{subequations}
\begin{multline}
F^{(\kappa)}(z)\equiv\sum_{n=1}^{\infty}V_n^{(\kappa)}\frac{z^n}{\sqrt{n}}
=V_0^{(\kappa)}\left[\frac{2}{b}\right]^{\frac{1}{2}}
\left[
-\frac{b}{4}-\Bigl(\Re F_c(\kappa)-\frac{b}{4}\Bigr)e^{-\kappa\arctan z}
\right.
\\
-\Bigl(\frac{\pi}{2\sqrt{3}}\frac{\nu-1}{\nu+1}
+\frac{2i}{\kappa}\Bigr)
-\log iz+2if^{(\kappa)}(z)
\\
\left.
-\mathrm{LerchPhi}(e^{-4i\arctan z},1,1+\frac{\kappa}{4i})
e^{-4i\arctan z}e^{-\kappa\arctan z}
\right].
\label{summary:gencont}
\end{multline}
The action of the operator $C'$ on the eigenvectors can be
obtained from the identity on generating functions
\begin{equation}F^{(\kappa)}(-z)=F^{(-\kappa)}(z)\quad\Longleftrightarrow\quad
CV^{(\kappa)}=V^{(-\kappa)}.
\end{equation}The norm of the vectors from continuous spectrum is given by
\begin{equation}(\!(V^{(\kappa)},\,V^{(\kappa')})\!)
=\frac{2}{b}|V_0^{(\kappa)}|^2\Nc(\kappa)\left[4+\kappa^2\Bigl(
\Re F_c(\kappa)-\frac{b}{4}
\Bigr)^2\right]\delta(\kappa-\kappa').
\label{norm:c}
\end{equation}\end{subequations}
Further we will assume that $V_0^{(\kappa)}$ is real and chosen in such way (see \eqref{V0:cont})
that the vectors from continuous spectrum \eqref{norm:c} are normalized
to $\delta$-function without any additional factors.
\end{itemize}

\medskip
\noindent \textbf{Partition of the unity.}\\
Since the operator $CM=\frac{1}{3}(1+CU+UC)$ is a hermitian
operator and vectors $V^{(\kappa)}$, $V^{(\xi)}$ and $V^{(\overline{\xi})}$
are the only its eigenvectors we conclude that these
vectors form a complete set. Therefore we can write the
following partition of the unit operator $\mathrm{Id}$
\begin{equation}\mathrm{Id}=\int_{-\infty}^{\infty}d\kappa\,V^{(\kappa)}\otimes V^{(\kappa)}
+V^{(\xi)}\otimes V^{(\xi)}
+V^{(\overline{\xi})}\otimes V^{(\overline{\xi})},
\label{unitypart}
\end{equation}where we assume that all vectors are normalized to $1$, i.e.
their zero components have been chosen as in equations \eqref{V0:cont}
and \eqref{V0:disc}.

\section{Spectrum of the operators defining ghost vertex}
\label{sec:spectrum-gh}
\setcounter{equation}{0}

\subsection{Spectrum of the operators $C\tilde{U}$ and $\tilde{U}C$ (ghosts)}
\label{sec:sp-gh:sp}
\begin{figure}[!t]
\centering
\includegraphics[width=290pt]{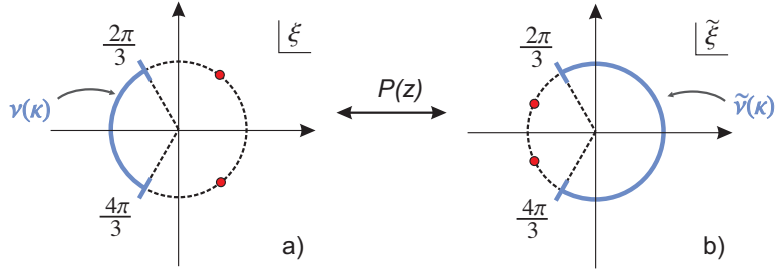}
\caption{On the pane a) we show the spectrum of the operator
$CU$, which defines the $\star$-product in the matter sector.
On the pane b) we show the spectrum of the operator $C\tilde{U}$,
which defines the $\star$-product in the ghost sector.
The arrow between the panes shows the relation among these two
operators.}
\label{fig:gh}
\end{figure}

In the section~\ref{sec:3string} we have shown
that the operator $C\tilde{U}$ for ghost sector
and the operator $CU$ for the matter sector are related
by $PSL(2,\Zh)$ transformation $P$ \eqref{P(z)}.
From this it follows that the eigenvectors of the operator $C\tilde{U}$
are the same as of those $CU$, and the eigenvalues are related
buy the transformation $P$. The resulting
spectrum of the operator $C\tilde{U}$
is presented on Figure~\ref{fig:gh}b).

So the spectrum of $C\tilde{U}$ consists of two branches: continuous
and discrete.
\begin{itemize}
\item \textbf{Discrete spectrum}\\
The discrete spectrum of operator $C\tilde{U}$ consist
of two points $\tilde{\xi}$ and $\overline{\tilde{\xi}}$, which
are defined as
\begin{equation}\tilde{\xi}\equiv P(\xi)=-\frac{2+\cosh\eta-i\sqrt{3}\sinh\eta}{1+2\cosh\eta}.
\end{equation}The eigenvectors corresponding to these eigenvalues
are $V^{(\tilde{\xi})}\equiv V^{(\xi)}$
and $V^{(\overline{\tilde{\xi}})}\equiv V^{(\overline{\xi})}$, and
their generating functions are given by
equation \eqref{summary:gendisc}.

\item \textbf{Continuous spectrum}
The discrete spectrum of operator $C\tilde{U}$ consists
of an arc shown on the Figure~\ref{fig:gh}. The parametrization
of this arc is given by function $\tilde{\nu}(\kappa)$
\begin{equation}\tilde{\nu}(\kappa)\equiv P(\nu(\kappa))=
-\frac{2-\cosh\frac{\pi\kappa}{2}+i\sqrt{3}\sinh\frac{\pi\kappa}{2}}{1-2\cosh\frac{\pi\kappa}{2}}.
\label{tilde-nu(kappa)}
\end{equation}The eigenvectors corresponding to these eigenvalues
are $V^{(\kappa)}$, and
their generating functions are given by
equation \eqref{summary:gencont}.
\end{itemize}

\noindent \textbf{Partition of the unity.}\\
Partition of the unit operator is the same as in \eqref{unitypart}.

\subsection{Spectrum of the operators $C'\tilde{U}'$ and $\tilde{U}'C'$ (ghosts)}
\label{sec:sp-gh:spU}
By analogy to the matter sector the operators $C\tilde{U}$ and $\tilde{U}C$
can be presented in the following block from
\begin{equation}C\tilde{U}=
\begin{pmatrix}
\tilde{U}_{00} & K^t\\
K & C'\tilde{U}'+\lambda K\otimes K^t
\end{pmatrix},
\label{gh:CU'}
\end{equation}where $\lambda$ is fixed by the  condition that $C\tilde{U}$ is
unitary operator, i.e.
\begin{equation}\lambda=\frac{1}{\tilde{U}_{00}\pm 1},\quad
\tilde{U'}\overline{K}=\mp \overline{K}\quad\text{and}\quad
\sum_{n=1}^{\infty}\overline{K}_nK_n=1-\tilde{U}_{00}^2.
\end{equation}Using the fact that operators $C\tilde{U}$ and $CU$ (for $b=2$) are related by
the transformation \eqref{P(z)} we can easily 
obtain the expressions for $\tilde{U}_{00}$, $\lambda$, $K$ and $C'\tilde{U}'$
\begin{subequations}
\begin{alignat}{2}
\tilde{U}_{00}&=\frac{1-\tilde{V}_{00}}{1+\tilde{V}_{00}},
\quad \tilde{V}_{00}=3\log 3
&\qquad
\lambda&=(1+\tilde{U}_{00})^{-1};
\\
K&=-\frac{3}{\sqrt{2}}(\tilde{U}_{00}+1)(C'U'+2)^{-1}\,W,
&\qquad
C'\tilde{U}'&=P(C'U').
\end{alignat}
\end{subequations}

Now we can conclude that the spectrum of the operator
$C'\tilde{U}'$ is only continuous and consists of
the eigenvectors defined by generating function \eqref{gen-func}
and corresponding eigenvalues are given by equation \eqref{tilde-nu(kappa)}.

\label{gh:discuss}
Basically only the right column of the operator \eqref{gh:CU'}
is appeared in the ghost vertex \eqref{V3:ghost}. Therefore
we need to know how represent this column in the diagonal form.
In principal it is possible to diagonilize the matrix $(CU)_{mn}$, $m,n\geqslant 1$.
To this end notice that the matrix $(CU)_{mn}$ is not a unitary one, i.e.
\begin{equation*}\sum_{n=1}^{\infty}(UC)_{mn}(CU)_{nk}=\delta_{nk}-\overline{K}_n\otimes K_{k}.
\end{equation*}The spectrum  of this operator consists of two branches continuous and
discrete. The eigenvalues from the continuous spectrum are represented by
the function $\tilde{\nu}(\kappa)$, and the corresponding eigenvectors
are complex (this is very different from the cases we had before).
There are two points in the discrete spectrum, one eigenvalue is
inside the unite circle and another one is outside. The corresponding
eigenvectors are also complex. Since all the eigenvectors
are complex it looks impossible to apply them to diagonalize the
vertex \eqref{gh:CU'}.

So will diagonalize only the operator $C'\tilde{U}'$ and express
the vector $K$ in terms of $v^{(\kappa)}$. The scalar product
of $K$ and $\overline{K}$ with $v^{(\kappa)}$ is given by
\begin{equation}(v^{(\kappa)},K)=(1+\tilde{U}_{00})\frac{1-\tilde{\nu}(\kappa)}{\kappa}
\quad\text{and}\quad
(v^{(\kappa)},\overline{K})=(1+\tilde{U}_{00})\frac{1-\overline{\tilde{\nu}}(\kappa)}{\kappa}.
\label{gh:vK}
\end{equation}

\section{Diagonal representation of the open string star}
\label{sec:diag}
\setcounter{equation}{0}


In this section we will rewrite the $3$-string vertex in the matter
and ghost sectors in the
oscillator basis related to the eigenvectors of operators
$CU$ and $UC$ ($C\tilde{U}$ and $\tilde{U}C$) found in the previous section.

\subsection{Matter $3$-vertex}
\label{sec:diag:m}
Let us introduce new oscillators
\begin{subequations}
\begin{equation}a_{\kappa}=\sum_{n=0}^{\infty}V_{n}^{(\kappa)}a_n,
\quad
a_{\xi}=\sum_{n=0}^{\infty}V_{n}^{(\xi)}a_n
\quad\text{and}\quad
a_{\overline{\xi}}=\sum_{n=0}^{\infty}V_{n}^{(\overline{\xi})}a_n.
\end{equation}Using the completeness of the basis \eqref{unitypart} we can
express old oscillators as
\begin{equation}a_n=\int_{-\infty}^{\infty} d\kappa\, V^{(\kappa)}_n a_{\kappa}
+V^{(\xi)}_n a_{\xi}+V^{(\overline{\xi})}_n a_{\overline{\xi}}.
\label{old-new}
\end{equation}The commutation relations for oscillators $a_{\kappa}$ are of the form
\begin{equation}[a_{\kappa,\mu},\,a_{\kappa',\nu}^{\dag}]=g_{\mu\nu}\delta(\kappa-\kappa')
\quad\text{and}\quad
[a_{\kappa,\mu},\,a_{\kappa',\nu}]=0.
\end{equation}The operator $C$ acts on the  on the new oscillators in the following way
\begin{equation}Ca_{\kappa}=a_{-\kappa}\quad\text{and}\quad Ca_{\xi}=a_{\overline{\xi}}.
\end{equation}\end{subequations}

Substitution of the expression \eqref{old-new} into \eqref{V3:matter} yields
the following representation for the matter $3$-string
vertex
\begin{equation}|V_3^m\rangle_{123}=\Nc_m^{-26}\exp\Bigl[
-\frac12\sum_{r,s} \int_{-\infty}^{\infty}d\kappa\,
\mu^{rs}(\kappa)(Ca_{\kappa}^{(r)\dag},a_{\kappa}^{(s)\dag})
-\sum_{r,s}
\mu^{rs}_{\xi}(Ca_{\xi}^{(r)\dag},a_{\xi}^{(s)\dag})
\Bigr]|\Omega_b\rangle_{123},
\label{prediag}
\end{equation}where
\begin{subequations}
\begin{align}
\mu^{rs}(\kappa)&=\frac13\left[
1+\alpha^{s-r}\nu(\kappa)+\alpha^{r-s}\overline{\nu}(\kappa)\right]
=\frac{1-2\delta_{r,s}+e^{\frac{\pi\kappa}{2}}\delta_{r+1,s}
+e^{-\frac{\pi\kappa}{2}}\delta_{r,s+1}}{1+2\cosh\frac{\pi\kappa}{2}},
\\
\mu^{rs}_{\xi}&=\frac13\left[
1+\alpha^{s-r}\xi+\alpha^{r-s}\overline{\xi}\right]
=\frac{1-2\delta_{r,s}-e^{\eta}\delta_{r+1,s}
-e^{-\eta}\delta_{r,s+1}}{1-2\cosh\eta}
.
\end{align}
\label{murs}
\end{subequations}
Equation \eqref{prediag} gives almost diagonal representation
of the interaction vertex. To write it in diagonal form
we need to introduce oscillator basis which has definite
parity with respect to the operator $C$. To this end
 let us introduce even and odd oscillator modes
\begin{subequations}
\begin{alignat}{4}
e_{\kappa}&=\frac{a_{\kappa}+Ca_{\kappa}}{\sqrt{2}},
&\qquad
o_{\kappa}&=\frac{a_{\kappa}-Ca_{\kappa}}{i\sqrt{2}},
& \qquad
e_{\xi}&=\frac{a_{\xi}+Ca_{\xi}}{\sqrt{2}},
&\qquad
o_{\xi}&=\frac{a_{\xi}-Ca_{\xi}}{i\sqrt{2}};
\\
a_{\kappa}&=\frac{e_{\kappa}+io_{\kappa}}{\sqrt{2}},
&
Ca_{\kappa}&=\frac{e_{\kappa}-io_{\kappa}}{\sqrt{2}},
&
a_{\xi}&=\frac{e_{\xi}+io_{\xi}}{\sqrt{2}},
&
Ca_{\xi}&=\frac{e_{\xi}-io_{\xi}}{\sqrt{2}}.
\label{a-odd-even}
\end{alignat}
\label{odd-even-osc}
\end{subequations}
We have introduced a factor $i$ in the definition of the odd oscillators so that
along with the even ones they satisfy the same BPZ conjugation property \cite{0202087}
\begin{equation}\mathrm{bpz}\,o_{\kappa}=-o_{\kappa}^{\dag}\quad\text{and}\quad
\mathrm{bpz}\,e_{\kappa}=-e_{\kappa}^{\dag}.
\end{equation}Substitution of \eqref{a-odd-even} into the expression
for $3$-string vertex \eqref{prediag} yields
\begin{multline}
|V_3\rangle=\Nc_m^{-26}\exp\left[-\frac{1}{4}\sum_{r,s}
\int_0^{\infty}d\kappa\,(\mu^{rs}+\mu^{sr})\Bigl\{(e^{(r)\dag}_{\kappa},e^{(s)\dag}_{\kappa})
+(o^{(r)\dag}_{\kappa},o^{(s)\dag}_{\kappa})\Bigr\}
\right.
\\
-\frac{i}{4}\sum_{r,s}
\int_0^{\infty}d\kappa\,(\mu^{rs}-\mu^{sr})\Bigl\{(e^{(r)\dag}_{\kappa},o^{(s)\dag}_{\kappa})
-(o^{(r)\dag}_{\kappa},e^{(s)\dag}_{\kappa})\Bigr\}
\\
-\frac{1}{4}\sum_{r,s}
(\mu^{rs}_{\xi}+\mu^{sr}_{\xi})\Bigl\{(e^{(r)\dag}_{\xi},e^{(s)\dag}_{\xi})
+(o^{(r)\dag}_{\xi},o^{(s)\dag}_{\xi})\Bigr\}
\\
\left.
-\frac{i}{4}\sum_{r,s}
(\mu^{rs}_{\xi}-\mu^{sr}_{\xi})\Bigl\{(e^{(r)\dag}_{\xi},o^{(s)\dag}_{\xi})
-(o^{(r)\dag}_{\xi},e^{(s)\dag}_{\xi})\Bigr\}
\right]|\Omega_b\rangle_{123}.
\label{V3mdiag}
\end{multline}
So this expression gives the diagonal representation of the full $3$-string
interaction vertex in the matter sector. Using this representation in the next section
we will show that it is encodes a family of even Moyal algebras \cite{0202087}.

Notice that for $\kappa=0$ the odd oscillator $o_{\kappa=0}$
is equal to zero and we are left only with the even oscillator $e_{\kappa=0}$
(see \eqref{F:kappa=0}).
To derive the meaning of this mode it is convenient to introduce coordinate
and momentum oscillators
\begin{equation*}
x_{\kappa}=\frac{i}{\sqrt{2}}(a_{\kappa}-a_{\kappa}^{\dag})
\quad\text{and}\quad p_{\kappa}=\frac{1}{\sqrt{2}}(a_{\kappa}+a_{\kappa}^{\dag}).
\end{equation*}
Using formulae \eqref{F:kappa=0} and \eqref{V0:cont} we can
write $x_{\kappa=0}$ and $p_{\kappa=0}$ in the following way
\begin{subequations}
\begin{align}
x_{\kappa=0}&=\frac{1}{2\sqrt{\pi\ap}}\left[
x_0+2\sqrt{\ap}\sum_{n=1}^{\infty} (-1)^n x_{2n}
\right]=\frac{1}{2\sqrt{\pi\ap}}\,X\Bigl(\frac{\pi}{2}\Bigr);
\label{kappa=0}
\\
p_{\kappa=0}&=\sqrt{\pi\ap}\left[
\frac{1}{2\pi}p_0+\frac{1}{\pi\sqrt{\ap}}\sum_{n=1}^{\infty} \frac{(-1)^n}{2n} p_{2n}
\right].
\end{align}
\label{commutingmode}
\end{subequations}
So we get that one of the variables is proportional to $X(\frac{\pi}{2})$.

\subsection{Ghost $3$-vertex}
\label{sec:diag:g}
Let us first rewrite the ghost vertex \eqref{V3:ghost} using the
representation \eqref{gh:CU'} for the ghost operators $CU$.
To this end we notice that
\begin{subequations}
\begin{align}
(C\widetilde{M}^{rs})_{0m}&=\frac{1}{3}\alpha^{s-r}K_m+\frac{1}{3}\alpha^{r-s}\overline{K}_m,
\\
(C\widetilde{M}^{rs})_{nm}&=
(C\widetilde{M}^{'rs})_{nm}+\frac{1}{3(1+\tilde{U}_{00})}
\Bigl[\alpha^{s-r}\overline{K}_n\otimes K_m+\alpha^{r-s}K_n\otimes\overline{K}_m\Bigr].
\end{align}
\end{subequations}
Substitution of these expressions into \eqref{V3:ghost} yields
\begin{multline}
|V_3^{\text{gh}}\rangle_{123}=\exp\left[-\frac{1}{3}\sum_{r,s=1}^{3}\sum_{m=1}^{\infty}
b_{0}^{(r)}\Bigl[\alpha^{s-r}K_m+\frac{1}{3}\alpha^{r-s}\overline{K}_m\Bigr]
\sqrt{m}c_{-m}^{(s)}
\right.
\\
-\sum_{r,s=1}^3\sum_{m,n=1}^{\infty}b_{-n}^{(r)}\frac{1}{\sqrt{n}}
(C\widetilde{M}^{'rs})_{nm}\sqrt{m}c_{-m}^{(s)}
\\
\left.
-\frac{1}{3(1+\tilde{U}_{00})}\sum_{r,s=1}^3\sum_{m,n=1}^{\infty}
b^{(r)}_{-n}\frac{1}{\sqrt{n}}
\Bigl[\alpha^{s-r}\overline{K}_n\otimes K_m+\alpha^{r-s}K_n\otimes\overline{K}_m\Bigr]
\sqrt{m}c^{(s)}_{-m}
\right]|+\rangle_{123},
\end{multline}
Now it is obvious that we have to introduce the following oscillators
\begin{subequations}
\begin{alignat}{3}
b_{\kappa}^{\dag}&=\tfrac{1}{\sqrt{\Nc(\kappa)}}\sum_{n=1}^{\infty}\frac{v_{n}^{(\kappa)}}{\sqrt{n}}\,b_{-n},
&\quad
b^{\dag}&=
\tfrac{1}{\sqrt{1-\tilde{U}_{00}^2}}\sum_{n=1}^{\infty}\frac{K_{n}}{\sqrt{n}}\,b_{-n}^{\dag},
&\quad
\overline{b}^{\dag}&=\tfrac{1}{\sqrt{1-\tilde{U}_{00}^2}}\sum_{n=1}^{\infty}\frac{\overline{K}_n}{\sqrt{n}}\,b_{-n};
\\
c_{\kappa}^{\dag}&=\tfrac{1}{\sqrt{\Nc(\kappa)}}\sum_{n=1}^{\infty}v_{n}^{(\kappa)}\sqrt{n}c_{-n},
&\quad
c^{\dag}&=\tfrac{1}{\sqrt{1-\tilde{U}_{00}^2}}\sum_{n=1}^{\infty}K_{n}\sqrt{n}\,c_{-n},
&\quad
\overline{c}^{\dag}&=\tfrac{1}{\sqrt{1-\tilde{U}_{00}^2}}\sum_{n=1}^{\infty}\overline{K}_n\sqrt{n}c_{-n}.
\end{alignat}
\end{subequations}
Using formulae \eqref{gh:vK} we can easily express
the oscillators $c^{\dag}$ and $\overline{c}^{\dag}$ through
 $c_{\kappa}^{\dag}$ (and the same for
$b^{\dag}$ and $\overline{b}^{\dag}$)
\begin{subequations}
\begin{align}
c^{\dag}=\int_{-\infty}^{\infty}
d\kappa\,&\lambda(\kappa)\,c^{\dag}_{\kappa}
\quad\text{and}\quad
\overline{c}^{\dag}=\int_{-\infty}^{\infty}
d\kappa\,\overline{\lambda}(\kappa)\,c^{\dag}_{\kappa},
\\
&\text{where}\quad
\lambda(\kappa)=[\tilde{V}_{00}\,\Nc(\kappa)]^{-\frac12}\,
\frac{1-\tilde{\nu}(\kappa)}{\kappa}.
\end{align}
\end{subequations}
Using the completeness \eqref{unity} of the basis $v^{(\kappa)}$ we can
express old oscillators as
\begin{equation}c_{-n}=\int_{-\infty}^{\infty} d\kappa\, v^{(\kappa)}_n c_{\kappa}^{\dag}
\quad\text{and}\quad
b_{-n}=\int_{-\infty}^{\infty} d\kappa\, v^{(\kappa)}_n b_{\kappa}^{\dag}.
\label{gh:old-new}
\end{equation}The commutation relations for oscillators $b_{\kappa}$, $c_{\kappa}$
$b$ and $c$  are of the form
\begin{equation}\{b_{\kappa},\,c_{\kappa'}^{\dag}\}=\delta(\kappa-\kappa'),\quad
\{b_{\kappa},\,c^{\dag}\}=\lambda(\kappa)
\quad\text{and}\quad
\{b_{\kappa},\,\overline{c}^{\dag}\}=\overline{\lambda}(\kappa).
\end{equation}The operator $C$ acts on the  on the new oscillators in the following way
\begin{equation}Cc^{\dag}_{\kappa}=-c^{\dag}_{-\kappa}\quad\text{and}\quad
Cc^{\dag}=\overline{c}^{\dag}
\end{equation}and the same for $b_{\kappa}$, $b$ and $\overline{b}$.
Substitution of the expression \eqref{gh:old-new} into \eqref{V3:ghost} yields
the following representation for the matter $3$-string
vertex
\begin{multline}
|V_3^{\text{gh}}\rangle_{123}=\exp\left[
-\frac{1}{3}(1-\tilde{U}_{00}^2)^{\frac12}\sum_{r,s}
b_0^{(r)\dag}(\alpha^{s-r}c^{(s)\dag}+\alpha^{r-s}\overline{c}^{\dag})
\right.
\\
+\sum_{r,s} \int_{0}^{\infty}d\kappa\,
\tilde{\mu}^{rs}(\kappa)\Bigl[b_{\kappa}^{(r)\dag} c_{-\kappa}^{(s)\dag}
+b_{\kappa}^{(s)\dag} c_{-\kappa}^{(r)\dag}\Bigr]
\\
\left.
-\frac{1-\tilde{U}_{00}}{3}\sum_{r,s}\Bigl(
\alpha^{s-r}\overline{b}^{(r)\dag}c^{(s)\dag}
+\alpha^{r-s}b^{(r)\dag}\overline{c}^{(s)\dag}
\Bigr)
\right]|+\rangle_{123},
\label{gh:prediag}
\end{multline}
where
\begin{equation}\tilde{\mu}^{rs}(\kappa)=\frac13\left[
1+\alpha^{s-r}\tilde{\nu}(\kappa)+\alpha^{r-s}\overline{\tilde{\nu}}(\kappa)\right]
=\frac{-1+2\delta_{r,s}-e^{\frac{\pi\kappa}{2}}\delta_{r+1,s}
-e^{-\frac{\pi\kappa}{2}}\delta_{r,s+1}}{1-2\cosh\frac{\pi\kappa}{2}},
\label{gh:murs}
\end{equation}The equation \eqref{gh:prediag} gives almost diagonal representation
of the interaction vertex. To write it in the diagonal form
one can introduce oscillator basis which has definite
parity with respect to the operator $C$.


\section{Moyal structure}\label{sec:moyal}
\setcounter{equation}{0}


The aim of this section is to identify Moyal algebras
appeared in the expression for the $3$-string vertex
in the matter sector.

\subsection{Oscillator vertex for the Moyal product}
Let us remind the expression for the vertex representation
of the Moyal product obtained in \cite{0202087}.
To this end we have to introduce the Moyal product on
functions and the function-state correspondence.

For our purpose it is sufficient to consider functions
on the two-dimensional real space $\Rh^2$. The Moyal
product is given by the following integral operator
\begin{subequations}
\begin{equation}(f\ast g)(x_1)=\int_{\Rh^2\times\Rh^2}d^2x_2 d^2x_3\,
K(x_1,x_2,x_3)f(x_2)g(x_3),
\end{equation}where the kernel $K(x_1,x_2,x_3)$ is given by
\begin{equation}K(x_1,x_2,x_3)=\frac{1}{\pi^2\det\Theta}\,\exp\left(
-2i\bigl[
x_1^{\alpha}(\Theta^{-1})_{\alpha\beta}x_2^{\beta}+
x_2^{\alpha}(\Theta^{-1})_{\alpha\beta}x_3^{\beta}+
x_3^{\alpha}(\Theta^{-1})_{\alpha\beta}x_1^{\beta}
\bigr]
\right),
\end{equation}indexes $\alpha,\beta=1,2$ and we choose the coordinate
system in which the antisymmetric tensor $\Theta^{\alpha\beta}$ has the
diagonal form
\begin{equation}\Theta^{\alpha\beta}=
\begin{pmatrix}
0 & \theta\\
-\theta & 0
\end{pmatrix}.
\end{equation}\end{subequations}

The function state correspondence is defined as
\begin{subequations}
\begin{align}
f\leftrightarrow |f\rangle&=\int d^2x\, f(x)|-x\rangle
\\
\intertext{and conjugated state is given by}
\langle f|&=\int d^2x\, \langle x|f(x),
\end{align}
\end{subequations}
where $|x\rangle$ are eigenvectors of the coordinate operator and
they normalization is given by \eqref{normxx'}.
To be able to introduce the vertex operator we need to rewrite
the state $|x\rangle$ in terms of creation/ annihilation operators.
The creation/annihilation operators $a^{\dag},b^{\dag}$ and $a,b$
are related to coordinate $x^{\alpha}$ and
momentum $p_{\alpha}$ operators by formulae
\begin{subequations}
\begin{equation}x_{1}=\frac{i}{\sqrt{2}}(a-a^{\dag}),\quad
p_{1}=\frac{1}{\sqrt{2}}(a+a^{\dag})\quad\text{and}\quad
x_{2}=\frac{i}{\sqrt{2}}(b-b^{\dag}),\quad
p_{2}=\frac{1}{\sqrt{2}}(b+b^{\dag}).
\label{xxppab}
\end{equation}The state $\langle x|$ can written trough the operators $a_{\alpha},a^{\dag}_{\alpha}$
(we assume $a_{\alpha}=(a,b)$)
in the following form
\begin{equation}\langle x| =\frac{1}{\sqrt{2\pi}}\,
\langle 0|\exp\Bigl[
-\frac{1}{2}(x, x)-i\sqrt{2}\,(a,x)
+\frac12(a, a)
\Bigr].
\end{equation}\end{subequations}
Expressions for the conjugated state and eigenstates of momentum
operator are presented in Appendix~\ref{app:conventions} (for $b\ap=2$, $d=2$).

Now we can write the Moyal product in terms of vertices
\begin{equation}|f\ast g\rangle_1=\bigr({}_2\langle f|\otimes{}_3\langle g|\bigl)\,|V_3(\theta)\rangle_{123},
\end{equation}where the vertex $|V_3(\theta)\rangle$ has the following form\footnote{
The other way to obtain this expression is to notice that in the momentum
representation the $3$-vertex corresponding to the Moyal product
is given by
\begin{multline*}
| V_3 \rangle_{123}  = \frac{1}{2\pi}\int dp^{(1)} dp^{(2)} dp^{(3)}
\, \delta^{(2)} (p^{(1)} + p^{(2)} + p^{(3)})
\\
\times\exp\Bigl[\frac{i}{6}\Theta^{\alpha\beta}
(p^{(1)}_{\alpha}p^{(2)}_{\beta}+p^{(2)}_{\alpha}p^{(3)}_{\beta}+p^{(3)}_{\alpha}p^{(1)}_{\beta})
\Bigr]
|p^{(1)}\rangle_{1}\otimes|p^{(2)}\rangle_{2}\otimes|p^{(3)}\rangle_{3}.
\end{multline*}
To obtain the expression \eqref{V3(theta)} one needs
to substitute states $|p\rangle$ from \eqref{cp-eigenstates} and integrate over
momenta.} \cite{0202087}
\begin{multline}
|V_3(\theta)\rangle_{123}=\frac{2}{3\sqrt{\pi}}\frac{1}{1+\frac{\theta^2}{12}}
\exp\left[-\frac{1}{2}\frac{\theta^2-4}{\theta^2+12}(a^{(1)\dag}a^{(1)\dag}
+b^{(1)\dag}b^{(1)\dag}+c.p.)\right.
\\
-\frac{8}{\theta^2+12}(a^{(1)\dag}a^{(2)\dag}
+b^{(1)\dag}b^{(2)\dag}+c.p.)
\\
\left.-\frac{4i\theta}{\theta^2+12}(a^{(1)\dag}b^{(2)\dag}
-b^{(1)\dag}a^{(2)\dag}+c.p.)\right]|0\rangle_{123}.
\label{V3(theta)}
\end{multline}
Notice that we use a bit different normalization of coordinate
and momentum eigenstates as compared to \cite{0202087}, but
the final form of the vertex \eqref{V3(theta)} is precisely the same.

\subsection{Identification of Moyal structures}
In this subsection we construct a  correspondence of the
open string $\star$-product to the Moyal product.
Basically we will single out two types of Moyal algebra (continuous and discrete) in
the expression \eqref{V3mdiag} for the $3$-string vertex.
We show that deformation parameter of the continuous Moyal algebra
is parametrized by $\kappa\in[0,\infty)$
and the one of discrete algebra is related to $\eta$.

Let us now try to identify  the vertex \eqref{V3(theta)},
which represent the Moyal product, in the diagonal matter $3$-string
vertex \eqref{V3mdiag}. To this end we need to identify
\begin{equation}\{a^{\dag}_{\mu},b^{\dag}_{\mu}\}\leftrightarrow \{e^{\dag}_{\kappa,\mu},o^{\dag}_{\kappa,\mu}\}
\quad\text{or}\quad
\{a^{\dag}_{\mu},b^{\dag}_{\mu}\}\leftrightarrow \{e^{\dag}_{\xi,\mu},o^{\dag}_{\xi,\mu}\}.
\end{equation}Here we add indexes $\mu$ to the the oscillators $a^{\dag},b^{\dag}$
as compared to \eqref{xxppab}. These just
mean that we have $26$ copies of isomorphic Moyal algebras\footnote{
More precisely the oscillators $a_{\alpha,\mu}$ are in the
space $T(\Rh^2)\otimes T(\Rh^{26})$. The tensor $\Theta^{\alpha\beta}$ acts
on the first index (i.e. $\alpha$), while metric in the string target space $g_{\mu\nu}$
acts on the second one.}.
These identifications requires the conditions
\begin{subequations}
\begin{alignat}{3}
\mu^{11}(\kappa)&=\frac{\theta^2_{\kappa}-4}{\theta^2_{\kappa}+12},&\qquad
\mu^{12}(\kappa)+\mu^{21}(\kappa)&=\frac{16}{\theta^2_{\kappa}+12}, &\qquad
\mu^{12}(\kappa)-\mu^{21}(\kappa)&=\frac{8\theta_{\kappa}}{\theta^2_{\kappa}+12};
\\
\mu^{11}_{\xi}&=\frac{\theta^2_{\xi}-4}{\theta^2_{\xi}+12},&
\mu^{12}_{\xi}+\mu^{21}_{\xi}&=\frac{16}{\theta^2_{\xi}+12}, &
\mu^{12}_{\xi}-\mu^{21}_{\xi}&=\frac{8\theta_{\xi}}{\theta^2_{\xi}+12}.
\end{alignat}
\end{subequations}
Using the explicit expressions \eqref{murs} we can easily solve these equations
and find that
\begin{equation}\theta_{\kappa}=2\tanh\frac{\pi\kappa}{4}
\quad\text{and}\quad
\theta_{\xi}=2\coth\frac{\eta(b)}{2},
\label{nc-param}
\end{equation}where the function $\eta(b)$ is defined a solution of the equation \eqref{eta(b)}
(see also Figure~\ref{fig:3}).
$\theta_{\kappa}$ and $\theta_{\xi}$ are non-commutative parameters
associated to the continuous and
discrete Moyal algebras respectively. Notice that for $\kappa=0$ there is
only one nonzero oscillator $e_{\kappa=0}$, and $\theta_{\kappa=0}$
 is also zero. Therefore for $\kappa=0$ the associated Moyal algebra
 is actually a commutative one.

For continuous Moyal algebra the non-commutativity parameter $\theta_{\kappa}$
is belongs to the interval $[0,2)$. For discrete Moyal
algebra the non-commutativity parameter $\theta_{\xi}$ lies in the interval $(2,\infty)$.
From the equations \eqref{nc-param} and \eqref{eta(b)} follows that
when $b\to \infty$ (i.e. $\ap\to 0$) the parameter $\theta_{\xi}$ goes to $2$.
It seems that in the low energy limit of the SFT action
we have some non-commuting variables.
The analysis of the generating function for eigenvectors $V^{(\xi)}$
shows that it goes to $0$ as $e^{-e^b}$ as $b\to \infty$ (see
also Figure~\ref{fig:4}). Therefore regardless of the fact that
$\theta_{\xi}$ is finite for $b\to \infty$ actually there are no
oscillator modes corresponding to this non-commutativity.


\section{Discussion}
\setcounter{equation}{0}

Here I would like to summarize the main results of this
paper:
\begin{enumerate}
\item We explicitly diagonalize the Neumann matrices defining
$3$-string vertex in the matter and ghost sectors. The spectrum
of these operators is given through the spectrum of the unitary
auxiliary operators $CU$ and $C\tilde{U}$ respectively.
\begin{itemize}
\item Detailed spectrum of the operator $CU$ from the matter sector
can be found in Section~\ref{sec:summary}. The spectrum
of the matter Neumann matrices have been also found by
Bo Feng, Yang-Hui He and Nicolas Moeller \cite{0202176}.
To compare the results let us remind the spectrum of the operator
$M^{11}$ \eqref{murs}. The spectrum consists of two branches: continuous and discrete.
The eigenvalues from continuous spectrum are in the interval $[-1/3,0)$.
Each eigenvalue in the interval $(-1/3,0)$ is doubly degenerated,
while for eigenvalue $-1/3$ corresponds only one eigenvector.
The discrete spectrum is doubly degenerated and consist of one point in the interval
$(0,1)$. The position of this point depends on the value of parameter $b$.
Basically, the results presented here and the ones obtained in \cite{0202176} are
the same,
except for the  point $-1/3$ from the continuous spectrum,
for which there is one more eigenvector in \cite{0202176}. The origin of this difference
is in improper normalization \cite{BoFeng}, \cite{0203175} of the eigenvectors from
continuous spectrum used in \cite{0202176}.

\item The spectrum of operator $C\tilde{U}$ from ghost sector is given in
Section~\ref{sec:spectrum-gh}.
\end{itemize}

\item Using the diagonal representation of the Neumann matrices in
the matter sector we identify the Moyal structures appearing
in the matter $3$-string vertex  (Section~\ref{sec:moyal}).
There are two types of Moyal algebras appearing here:
continuous (found in \cite{0202087}) and discrete. The deformation parameter
of the continuous Moyal algebra is given by continuous function
$\theta(\kappa)$ defined in \eqref{nc-param}. The value of the continuous
deformation parameter
is restricted to the interval $[0,2)$.
The deformation parameter of the discrete Moyal algebra is given
by $\theta_{\xi}$ defined in \eqref{nc-param}. The value of the
discrete deformation parameter depends on the parameter $b$
and is in the interval $(2,\infty)$. As $b$ goes to infinity (i.e. $\ap\to 0$)
$\theta_{\xi}$ goes to $2$. The corresponding non-commutative modes
goes to zero as $e^{-e^b}$. Despite of the fact that $\theta_{\xi}=2$
for $\ap=0$ the non-noncommutative modes are equal to zero.

\item Notice that the commuting mode for the full string product
 is $X(\frac{\pi}{2})$ (see \eqref{commutingmode}),
while for the product reduced to the zero momentum sector
it is left momentum $P_L$ \cite{0202087}.
\end{enumerate}

Below we list some open questions:
\begin{enumerate}
\item Identification of ghost vertex algebra with the known ones
\cite{BK};

\item In the framework of \cite{0202087} try to establish
the exact correspondence between Witten's string product and Moyal product
\cite{BK};

\item Rewrite the BRST charge in the diagonal basis;

\item Investigate the sliver in the diagonal basis. Classify
all projectors of the open string field algebra.

\end{enumerate}


\section*{Acknowledgments}
I would like to thank H.~Liu, G.~Moore and B.~Zwiebach for useful discussions,
B.~Feng for correspondence on the spectrum of operator $M^{11}$,
A.~Konechny for discussion of the ghost vertices and comments
on the manuscript.
And especially I would like to thank S.~Lukyanov
for communication at various stages of my work.
I would like to acknowledge the hospitality of Lawrence Berkeley
National Laboratory, where
the final pages of the paper were written.
The work was supported in part by RFBR grant 02-01-00695.

\clearpage
\appendix
\section*{Appendix}
\addcontentsline{toc}{section}{Appendix}
\renewcommand {\theequation}{\thesection.\arabic{equation}}

\section{Normalization of the coordinate and momentum eigenstates}
\label{app:conventions}
\setcounter{equation}{0}


The momentum and coordinate eigenstates are given by
the following formulae
\begin{subequations}
\begin{align}
|p\rangle &=\left[\frac{2\pi}{b\ap}\right]^{-\frac{d}{4}}
\exp\Bigl[
-\frac{b\ap}{4}(p, p)+\sqrt{b\ap}(a_0^{\dag}, p)
-\frac12(a_0^{\dag}, a_0^{\dag})
\Bigr]
|\Omega_b\rangle,
\\
|x\rangle &=\frac{1}{(2\pi b\ap)^{d/4}}\,
\exp\Bigl[
-\frac{1}{b\ap}(x, x)+\frac{2i}{\sqrt{b\ap}}\,(a_{0}^{\dag},x)
+\frac12(a_0^{\dag}, a_0^{\dag})
\Bigr]
|\Omega_b\rangle.
\end{align}
\label{cp-eigenstates}
\end{subequations}
The scalar product $(\cdot,\cdot)$ and the vacuum state $|\Omega_b\rangle$
are defined by
\begin{equation}(p,p)\equiv g^{\mu\nu}p_{\mu}p_{\nu},
\quad (x,x)\equiv g_{\mu\nu}x^{\mu}x^{\nu},
\quad (x,p)\equiv x^{\mu}p_{\mu}
\quad\text{and}\quad a_n|\Omega_b\rangle=0\quad\text{for}\quad
n\geqslant 0,
\end{equation}where $g_{\mu\nu}$ is constant symmetric non-degenerate $d\times d$ matrix
(constant metric)
and $g^{\mu\nu}$ is its inverse.
The conjugated states to the eigenstates \eqref{cp-eigenstates}
are given by
\begin{subequations}
\begin{align}
\langle p|&=\left[\frac{2\pi}{b\ap}\right]^{-\frac{d}{4}}\langle\Omega_b|
\exp\Bigl[
-\frac{b\ap}{4}(p, p)-\sqrt{b\ap}(a_0, p)
-\frac12(a_0, a_0)
\Bigr]
\\
\langle x| &=\frac{1}{(2\pi b\ap)^{d/4}}\,
\langle\Omega_b|\exp\Bigl[
-\frac{1}{b\ap}(x, x)-\frac{2i}{\sqrt{b\ap}}\,(a_{0},x)
+\frac12(a_0, a_0)
\Bigr].
\end{align}
\label{eigenstates-cp}
\end{subequations}

\vspace{-1cm}
\begin{subequations}
\begin{align}
\intertext{Normalization of the momentum and coordinate eigensates:}
\langle p|p'\rangle&=\sqrt{\det g}\,\delta^{(d)}(p_{\mu}+p'{}_{\mu})
\quad\text{and}\quad
\langle x|x'\rangle=\frac{1}{\sqrt{\det g}}\,\delta^{(d)}(x^{\mu}+x'{}^{\mu});
\label{normxx'}
\\
\intertext{normalization of the function $\langle p | x\rangle$:}
\langle p|x\rangle &=\overline{\langle x|p\rangle}
 =\frac{1}{(2\pi)^{\frac{d}{2}}}\,e^{ip_{\mu}x^{\mu}};
\\
\intertext{string field:}
|t\rangle=&\int \frac{d^dp}{\sqrt{\det g}}\,t(p)|p\rangle,
\qquad t(-p)=\langle p|t\rangle;
\\
\intertext{Fourier transform:}
t(x)&=\frac{1}{(2\pi)^{\frac{d}{2}}}\int \frac{d^dp}{\sqrt{\det g}}\, e^{-ip_{\mu}x^{\mu}}t(p)
\quad\text{and}\quad
t(p)=\frac{1}{(2\pi)^{\frac{d}{2}}}\int d^dx\sqrt{\det g}\, e^{ip_{\mu}x^{\mu}} t(x).
\end{align}
\label{norm:norm}
\end{subequations}

\section{Conformal Ghosts}\label{app:spectrum}
\setcounter{equation}{0}

\subsection{Search for the discrete spectrum}
\subsubsection{Parametrization of the eigenvalues}
First of all, let us express the parameter $\kappa$ appearing in
the \eqref{nu(kappa)} through $\nu$. So we want to
solve the equation
\begin{equation*}(2\nu+1)\cosh\frac{\pi\kappa}{2}+2+\nu=-i\sqrt{3}\sinh\frac{\pi\kappa}{2}.
\end{equation*}This equation has two solutions:
\begin{subequations}
\begin{enumerate}
\item $\nu$-independent solution
\begin{equation}\cosh\frac{\pi\kappa}{2}=-\frac{1}{2}\quad\text{and}\quad
\sinh\frac{\pi\kappa}{2}=i\frac{\sqrt{3}}{2}\quad\Rightarrow\quad
\frac{\pi\kappa}{2}=\frac{2\pi i}{3}+2\pi i n.
\end{equation}
\item $\nu$-dependent solution
\begin{equation}\cosh\frac{\pi\kappa}{2}=-\frac{1}{2}\frac{\nu^2+4\nu+1}{\nu^2+\nu+1}\quad\text{and}\quad
\sinh\frac{\pi\kappa}{2}=-i\frac{\sqrt{3}}{2}\frac{\nu^2-1}{\nu^2+\nu+1}.
\end{equation}We will always assume that $\nu$ lies on the unit circle (see Figure~\ref{fig:1}),
i.e. $\nu=e^{i\phi}$.
It convenient to rewrite the formulae above in the following form
\begin{equation}\cosh\frac{\pi\kappa}{2}=-\frac{\cos\phi+2}{2\cos\phi+1}\quad\text{and}\quad
\sinh\frac{\pi\kappa}{2}=\sqrt{3}\frac{\sin\phi}{2\cos\phi+1}.
\end{equation}If $\frac{2\pi}{3}\leqslant\phi\leqslant\frac{4\pi}{3}$ then we will assume
$\frac{\pi\kappa}{2}\in \Rh$,
otherwise if $-\frac{2\pi}{3}\leqslant\phi\leqslant\frac{2\pi}{3}$
we will assume $\frac{\pi\kappa}{2}\in\Rh +\pi i$. Moreover it follows from
this parametrization that if $\kappa$ is a solution
for $\nu$ then $-\kappa$ is a solution for $\overline{\nu}$.
\end{enumerate}
\label{sol-kappa}
\end{subequations}

\subsubsection{Computation of the integral}
Now we are ready to compute the following integral
\begin{equation}(\overline{W},\frac{1}{C'U'-\xi}W)=\int_{-\infty}^{\infty}d\kappa
\frac{1}{\Nc(\kappa)}\frac{[(v^{(\kappa)},\,W)]^2}{\nu(\kappa)-\xi}
\stackrel{\ref{vWscalar}}{=}
2\int_{-\infty}^{\infty}\frac{d\kappa}{\kappa}\,
\frac{[2+\nu(\kappa)+\overline{\nu}(\kappa)]}{\kappa\Nc(\kappa)}\,
\frac{\nu(\kappa)}{\nu(\kappa)-\xi}
.
\label{WoW0}
\end{equation}Substitution of the expressions \eqref{nu(kappa)} and \eqref{N(kappa)}
and some simplification yields
\begin{equation}(W,\frac{1}{C'U'-\xi}W)=2\int_{-\infty}^{\infty}\frac{dx}{x}\,
\frac{\sinh\frac{x}{2}}{\cosh\frac{x}{2}}\frac{1}{(1+2\cosh x)}\,
\frac{2+\cosh x+i\sqrt{3}\sinh x}{(2\xi+1)\cosh x+i\sqrt{3}\sinh x
+2+\xi},
\label{WxiW}
\end{equation}where $x=\frac{\pi\kappa}{2}$.
The analysis performed in \eqref{sol-kappa} yields to the
following parametrization of the $\xi$:
\begin{equation}\cosh(\eta+i\pi)=-\frac{1}{2}\frac{\xi^2+4\xi+1}{\xi^2+\xi+1}\quad\text{and}\quad
\sinh(\eta+i\pi)=-i\frac{\sqrt{3}}{2}\frac{\xi^2-1}{\xi^2+\xi+1}.
\label{xi-param}
\end{equation}\begin{figure}[!t]
\centering
\includegraphics[width=320pt]{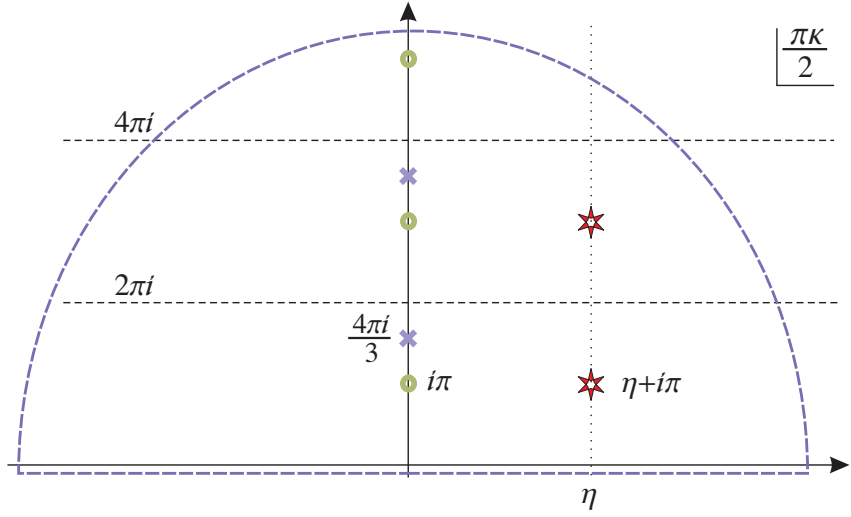}
\caption{Contour of integration for expression \eqref{WxiW}.}
\label{fig2}
\end{figure}
The integral \eqref{WxiW} has the following poles (see Figure~\ref{fig2}):
\begin{itemize}
\item $x=\pi i+2\pi i n$, $n\geqslant 0$;
\item $x=\eta+i\pi+2\pi in$, $n\geqslant 0$;
\item $x=\frac{4\pi i}{3}+2\pi i n$, $n\geqslant 0$.
\end{itemize}
One sees that we have two cases $\eta=0$ and
$\eta\ne 0$.

Let us first consider the case $\eta\ne 0$.
The computation of the residues in the upper half plane leads to the following result
\begin{multline}
(\overline{W},\frac{1}{C'U'-\xi}W)=2\sum_{n=0}^{\infty}\left[
\frac{2}{\xi-1}\frac{1}{n+\frac{1}{2}}+\frac{1}{n+\frac{2}{3}}
-\frac{\xi+1}{\xi-1}\frac{1}{n+\frac{1}{2}+\frac{\eta}{2\pi i}}
\right]
\\
=
\frac{\xi+1}{\xi-1}\,2F(\eta)+2\bigl[\psi(\tfrac{1}{2})-\psi(\tfrac{2}{3})\bigr],
\label{WoW1}
\end{multline}
where $\psi(x)$ is the logarithmic derivative of the Euler $\Gamma$-function
\begin{equation*}\begin{split}
\psi(x)&=-\gamma_E-\sum_{n=0}^{\infty}\Bigl(\frac{1}{n+x}-\frac{1}{n+1}\Bigr),
\\
\psi(\tfrac{1}{2})&=-\gamma_E-\log 4
\quad\text{and}\quad
\psi(\tfrac{2}{3})=-\gamma_E+\frac{\pi}{2\sqrt{3}}-\log 3\sqrt{3}
\end{split}
\end{equation*}and
\begin{equation}F(\eta)=\psi(\tfrac{1}{2}+\tfrac{\eta}{2\pi i})-\psi(\tfrac{1}{2}).
\label{F(eta)}
\end{equation}
If $\eta=0$ we need only to know that the integral \eqref{WxiW} is finite. It follows from
equation \eqref{EQ-desc} that $\eta=0$ and $\xi=1$ is
its solution only for $b=0$.

\subsubsection{Solving of the equation}
\begin{figure}
\centering
\includegraphics[width=500pt]{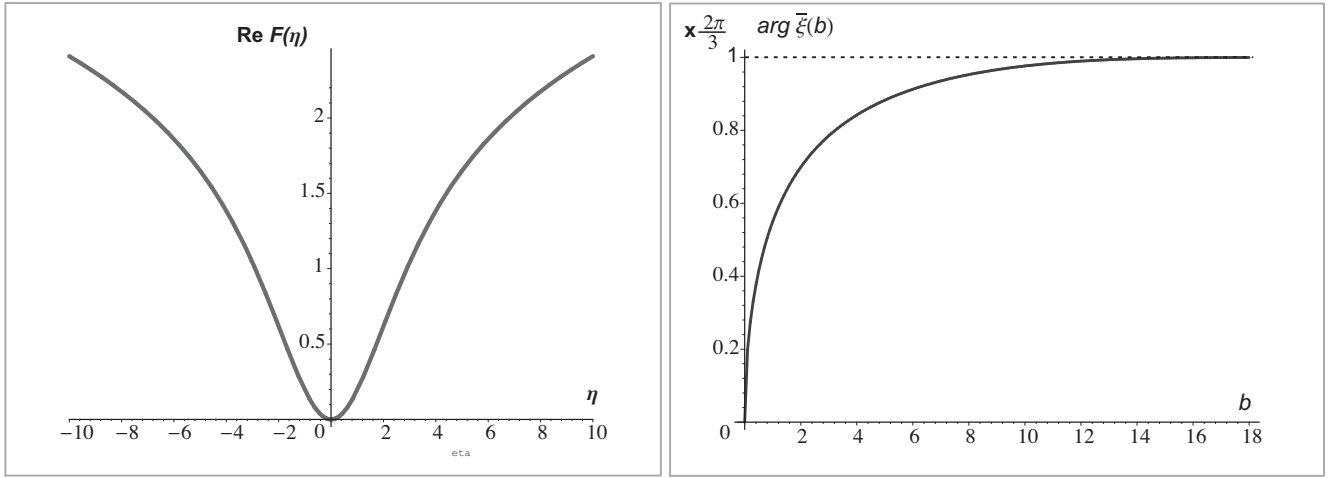}
\caption{On the left pane we present the graph of function $\Re F(\eta)$. As $\eta$
goes to $\infty$ this function grows logarithmically. On the right pane
we present the graph of $\arg \bar{\xi}(b)$. As $b$ grows this function goes to its maximal
value $\frac{2\pi}{3}$ exponentially fast.}
\label{fig:3}
\end{figure}

Let us simplify expression \eqref{WoW1}. To this end we
substitute $\psi(\frac12)$ and $\psi(\frac{2}{3})$ and
notice that\footnote{This expression follows
from the following formula for the $\Gamma$-function
$$
\Gamma(1-z)\Gamma(z)=\frac{\pi}{\sin \pi z}.
$$}
\begin{equation}\Im F(\eta)=-\frac{\pi}{2}\tanh\frac{\eta}{2}\equiv
\frac{\pi}{2i\sqrt{3}}\frac{\xi-1}{\xi+1}.
\label{psi1}
\end{equation}Therefore \eqref{WoW1} takes a form
\begin{equation}(\overline{W},\frac{1}{C'U'-\xi}W)=V_{00}+\frac{\xi+1}{\xi-1}\,2\Re F(\eta).
\label{WoW2}
\end{equation}Substitution of this expression into equation
\eqref{EQ-desc} yields the equation
\begin{equation}\Re F(\eta)=\frac{b}{4}.
\label{psi2}
\end{equation}$\Re F(\eta)$ is an even function with $F(0)=0$ (see Figure~\ref{fig:3}),
therefore for each $b>0$ this equation has two solutions $\pm\eta(b)$.
On Figure~\ref{fig:3}
we presented the $\arg\bar{\xi}(b)$ for this solution.

\subsection{Generating function for discrete spectrum}
In this section we will construct the generating function
for the vector defined by \eqref{EQ-Vdisc}. To this end
we rewrite the expression \eqref{EQ-Vdisc} for the components
of the vector $V^{(\xi)}$ in the following
form
\begin{multline*}
V_{m}^{(\xi)}=\frac{V_0^{(\xi)}}{\sqrt{b}}(\xi-1)
\left(\frac{1}{C'U'-\xi}\right)_{mn}W_n
\stackrel{\ref{unity}}=\frac{V_0^{(\xi)}}{\sqrt{b}}(\xi-1)\int_{-\infty}^{\infty}
\frac{d\kappa}{\Nc(\kappa)}\, v_m^{(\kappa)} \Bigl(v^{(\kappa)},\,
\frac{1}{C'U'-\xi}W\Bigr)
\\
\stackrel{\ref{vWscalar}}{=}
-\frac{V_0^{(\xi)}}{\sqrt{b}}(\xi-1)\int_{-\infty}^{\infty}
\frac{d\kappa}{\kappa\Nc(\kappa)}\,v_m^{(\kappa)}
\frac{\sqrt{2}[1+\nu(\kappa)]}{\nu(\kappa)-\xi}.
\end{multline*}
The generating function for the vector $V^{(\xi)}_m$ is defined as
\begin{equation}F^{(\xi)}(z)=\sum_{m=1}^{\infty}V_{m}^{(\xi)}\frac{z^m}{\sqrt{m}}=
-\sqrt{2}\frac{V_0^{(\xi)}}{\sqrt{b}}(\xi-1)\int_{-\infty}^{\infty}
\frac{d\kappa}{\kappa\Nc(\kappa)}\,\frac{f^{(\kappa)}(z)[1+\nu(\kappa)]}{\nu(\kappa)-\xi}.
\label{gen-func-xi1}
\end{equation}
To get the explicit expression for this generating function
we need to compute the integral
\begin{equation*}
\int_{-\infty}^{\infty}
\frac{d\kappa}{\kappa\Nc(\kappa)}\,\frac{f^{(\kappa)}(z)[1+\nu(\kappa)]}{\nu(\kappa)-\xi}
\stackrel{\ref{nu(kappa)}}{=}\int_{-\infty}^{\infty}
\frac{d x}{x}\,\frac{1-e^{-x\frac{2}{\pi}\arctan z}}{2\sinh x}
\frac{1-\cosh x+i\sqrt{3}\sinh x}{(2\xi+1)\cosh x+i\sqrt{3}\sinh x+2+\xi}
\end{equation*}
We want to compute this integral by summing residues in the UHP.
To this end we add $+ix\delta$ ($\delta>0$) to the exponent.
This integral has the following poles
\begin{itemize}
\item $x=\pi i+2\pi i n$, $n\geqslant 0$;
\item $x=\eta+\pi i+2\pi i n$, $n\geqslant 0$.
\end{itemize}
Summing over these residues yields
\begin{multline}
=-\frac{1}{\xi-1}\sum_{n=0}^{\infty}\left[
-\frac{1}{n+\frac12}+\frac{e^{-4i(n+\frac12)(\arctan z -i\delta)}}{n+\frac12}
+\frac{1}{n+\frac12+\frac{\eta}{2\pi i}}
-\frac{e^{-4i(n+\frac12+\frac{\eta}{2\pi i})(\arctan z -i\delta)}}{n+\frac12+\frac{\eta}{2\pi i}}
\right]
\\
=\frac{1}{\xi-1}\left[
\psi(\tfrac12+\tfrac{\eta}{2\pi i})-\psi(\tfrac12)
+e^{-2i(1+\frac{\eta}{\pi i})\arctan z}
\mathrm{LerchPhi}(e^{-4i\arctan z},1,\frac12+\frac{\eta}{2\pi i})
\right.
\\
\left.
-e^{-2i\arctan z}
\mathrm{LerchPhi}(e^{-4i\arctan z},1,\frac12)
\right],
\end{multline}
where
$$
\mathrm{LerchPhi}(w,1,x)=
\sum_{n=0}^{\infty}\frac{w^n}{n+x}
=
\frac{1}{x}\,{}_2F_1(1,x;x+1;w).
$$
Using the fact that
$${}_2F_1(1,\frac12;\frac32;w)=
\frac{1}{\sqrt{w}}\arctanh\sqrt{w},
\qquad
\arctanh e^{-4i\arctan z}=-\frac12\log iz
$$
and the expressions
\eqref{psi1}, \eqref{psi2} we can write the following
formula for the generating function $F^{(\xi)}(z)$
\begin{multline}
F^{(\xi)}(z)=-\sqrt{\frac{2}{b}}\,V_0^{(\xi)}\left[
\frac{b}{4}+\frac{\pi}{2\sqrt{3}}\frac{\xi-1}{\xi+1}
+\log iz
\right.
\\
\left.
+e^{-2i(1+\frac{\eta}{\pi i})\arctan z}
\mathrm{LerchPhi}\bigl(e^{-4i\arctan z},1,\tfrac12+\tfrac{\eta}{2\pi i}\bigr)
\right],
\end{multline}
where $F(\eta)$ is defined by \eqref{F(eta)} and $\eta$ is
related to $\xi$ by formula \eqref{xi-param}.

Having these formulae one can easily calculate how complex conjugation
and operator $C$ change the eigenvectors:
\begin{equation}F^{(\xi)}(-z)\equiv F^{(\overline{\xi})}(z)\quad
\text{and}\quad
\overline{F^{(\xi)}}(z)\equiv F^{(\xi)}(z)\quad\text{iff}\quad
V_0^{(\xi)}\in \Rh.
\end{equation}
\subsection{Normalization of the vectors
from the discrete spectrum}
\begin{figure}[!t]
\centering
\includegraphics[width=220pt]{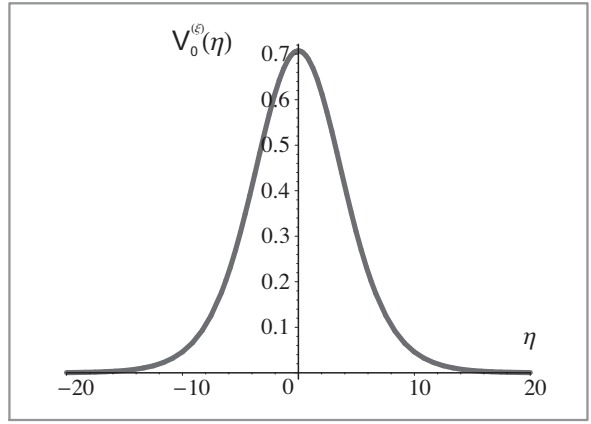}
\caption{Plot of the zero component $V_0^{(\xi)}$ of the vectors
from continuous spectra. If we choose zero component in this way
the vectors will be normalized to $1$.}
\label{fig:4}
\end{figure}

In this subsection we will compute the following scalar product
\begin{equation}(\!(V^{(\xi)},V^{(\xi)})\!):=|V^{(\xi)}_0|^2+(V^{(\xi)},V^{(\xi)}).
\label{Vnorm}
\end{equation}The expression \eqref{Vnorm} yields
\begin{multline}
(V^{(\xi)},V^{(\xi)})\stackrel{\ref{EQ-Vdisc}}{=}
\frac{|V^{(\xi)}_0|^2}{b}|\xi-1|^2
\Bigl(W,\frac{1}{U'C'-\overline{\xi}}\frac{1}{C'U'-\xi}W\Bigr)
\\
\stackrel{\ref{unity},\ref{vWscalar}}{=}\frac{|V^{(\xi)}_0|^2}{b}|\xi-1|^2
2\int_{-\infty}^{\infty}\frac{d\kappa}{\kappa}
\frac{2+\nu+\overline{\nu}}{\kappa\Nc(\kappa)}
\frac{1}{(\nu-\xi)(\overline{\nu}-\overline{\xi})}
\\
=
\frac{|V^{(\xi)}_0|^2}{b}|\xi-1|^2
(-\xi)\frac{\pd}{\pd\xi}
\left[2\int_{-\infty}^{\infty}\frac{d\kappa}{\kappa}
\frac{2+\nu+\overline{\nu}}{\kappa\Nc(\kappa)}
\frac{\nu}{\nu-\xi}\right]
\\
\stackrel{\ref{WoW0},\ref{WoW2}}{=}
\frac{|V^{(\xi)}_0|^2}{b}|\xi-1|^2
\left[
-\frac{b}{|\xi-1|^2}+\frac{2\sqrt{3}}{i}\frac{\xi+1}{\xi-1}\frac{1}{\xi+\overline{\xi}+1}
\frac{\pd}{\pd\eta}\Re F(\eta)
\right].
\end{multline}
Substitution of this expression into \eqref{Vnorm}
and simplification by use of \eqref{psi2} yields
\begin{equation}(\!(V^{(\xi)},V^{(\xi)})\!)=|V_0^{(\xi)}|^2\sinh\eta\, \frac{\pd}{\pd\eta}
\Bigl[\log \Re F(\eta)\Bigr].
\end{equation}The function on the rhs is positive and more or equal to $2$.
So to have the norm equal to one, we have
to choose $V_0^{(\xi)}$ equal to
\begin{equation}V_{0}^{(\xi)}=\left(\sinh\eta\, \frac{\pd}{\pd\eta}
\Bigl[\log \Re F(\eta)\Bigr]\right)^{-1/2}.
\label{V0:disc}
\end{equation}The graphic of this function is presented
on the Figure~\ref{fig:4}.
Notice that as $\eta\to\infty$ $V^{(\xi)}_0$ goes to zero
exponentially fast.

\subsection{Generating function for the continuous spectrum}
\subsubsection{Calculation of $A^{(\kappa)}$}
First of all we need to obtain the expression
for the $A^{(\kappa)}$ defined by
equation \eqref{cont:A}. To this end we have to
calculate the integral
\begin{multline}
\left(\overline{W},\,\mathscr{P}\frac{1}{C'U'-\nu(\kappa)}W\right)
=2\int_{-\infty}^{\infty}\frac{d\kappa'}{\kappa'}\frac{2+\nu+\overline{\nu}}{\kappa \Nc(\kappa')}
\,\mathscr{P}\frac{\nu(\kappa')}{\nu(\kappa')-\nu(\kappa)}
\\
=2\int\frac{dx}{x}\frac{\sinh\frac{x}{2}}{\cosh\frac{x}{2}}
\frac{1}{1+2\cosh x}\,\mathscr{P}\frac{2+\cosh x+i\sqrt{3}\sinh x}{(2\nu+1)\cosh x
+i\sqrt{3}\sinh x+2+\nu}.
\end{multline}
This integral has the following poles:
\begin{itemize}
\item $x=\pi i+2\pi i n$, $n\geqslant 0$;
\item $x=\frac{\pi\kappa}{2}+2\pi i (n+1)$, $n\geqslant 0$;
\item $x=\frac{\pi\kappa}{2}$, one half of the residue;
\item $x=\frac{4\pi i}{3} i+2\pi i n$, $n\geqslant 0$.
\end{itemize}
Calculation of the residues yields
\begin{multline}
=2\sum_{n=0}^{\infty}\left[
\frac{2}{\nu-1}\frac{1}{n+\frac12}-\frac{\nu+1}{\nu-1}\frac{1}{n+1+\frac{\kappa}{4i}}
+\frac{1}{n+\frac{2}{3}}
\right]
-\frac{\nu+1}{\nu-1}\frac{4i}{\kappa}
\\
=-\frac{\pi}{\sqrt{3}}+V_{00}+2\frac{\nu+1}{\nu-1}
\left[\psi(1+\tfrac{\kappa}{4i})-\psi(\tfrac12)-\frac{2i}{\kappa}\right]
\\
=V_{00}+\frac{\nu+1}{\nu-1}\,2\Re F_c(\kappa),
\label{cont:int:A}
\end{multline}
where
\begin{equation}F_c(\kappa)=\psi(1+\tfrac{\kappa}{4i})-\psi(\tfrac{1}{2})
\end{equation}and to obtain the last line in the \eqref{cont:int:A}
we have used the fact that
\begin{equation*}\Im F_c(\kappa)=\frac{2}{\kappa}-\frac{\pi}{2}\frac{\sinh\frac{\pi\kappa}{2}}
{\cosh\frac{\pi\kappa}{2}}\equiv
\frac{2}{\kappa}-\frac{\pi i}{2\sqrt{3}}\frac{\nu-1}{\nu+1}.
\end{equation*}Substitution of \eqref{cont:int:A} into \eqref{cont:A} yields
the following expression for $A^{(\kappa)}$:
\begin{equation}A^{(\kappa)}=V_0^{(\kappa)}\sqrt{\frac{2}{b}}\,\kappa\Bigl(
\Re F_c(\kappa)-\frac{b}{4}
\Bigr).
\label{A(kappa)}
\end{equation}
\subsubsection{Generating function}
The generating function for the vector $V^{(\kappa)}_m$ is
defined in the following way
\begin{multline}
F^{(\kappa)}(z)=\sum_{m=1}^{\infty}V_{m}^{(\kappa)}\frac{z^m}{\sqrt{m}}
\\
\stackrel{\ref{cont:V}}{=}
A^{(\kappa)}f^{(\kappa)}(z)
-[1-\nu(\kappa)]
\frac{V_{0}^{(\kappa)}}{\sqrt{b}}
\,\int \frac{d\kappa'}{\Nc(\kappa')}\,
f^{(\kappa')}(z)\left(v^{(\kappa')},\,\mathscr{P}\frac{1}{CU-\nu(\kappa)}W
\right).
\label{gen-func-kappa}
\end{multline}
Let us start from the computing the integral in the rhs:
\begin{multline}
\int \frac{d\kappa'}{\Nc(\kappa')}\,
f^{(\kappa')}(z)\mathscr{P}\frac{(v^{(\kappa')},\,W)}{\nu(\kappa')-\nu(\kappa)}=
-\sqrt{2}\int \frac{d\kappa'}{\kappa'\Nc(\kappa')}\,
\frac{1}{\kappa'}(1-e^{-\kappa\arctan z})
\mathscr{P}\frac{1+\nu(\kappa')}{\nu(\kappa')-\nu(\kappa)}
\\
=-\sqrt{2}\int_{-\infty}^{\infty}\frac{dx}{x}
\frac{1-e^{-x\frac{2}{\pi}\arctan z}}{2\sinh x}
\,\mathscr{P}\frac{1-\cosh x+i\sqrt{3}\sinh x}{(2\nu+1)\cosh x
+i\sqrt{3}\sinh x+2+\nu}.
\end{multline}
This integral has the following poles
\begin{itemize}
\item $x=\pi i+2\pi n$, $n\geqslant 0$;
\item $x=\frac{\pi\kappa}{2}+2\pi (n+1)$, $n\geqslant 0$;
\item $x=\frac{\pi\kappa}{2}$, half of the pole.
\end{itemize}
Calculation of the residues yields
\begin{multline}
=-\frac{\sqrt{2}}{1-\nu}\sum_{n=0}^{\infty}
\left[
-\frac{1-e^{-(n+\frac12)4i\arctan z}}{n+\frac12}
+\frac{1-e^{-(n+1+\frac{\kappa}{4i})4i\arctan z}}{n+1+\frac{\kappa}{4i}}
\right]
-\frac{\sqrt{2}}{1-\nu}\frac{2i}{\kappa}(1-e^{-\kappa\arctan z})
\\
=-\frac{\sqrt{2}}{1-\nu}
\Bigl[
\psi(\tfrac12)-\psi(1+\tfrac{\kappa}{4i})+\frac{2i}{\kappa}(1-e^{-\kappa\arctan z})
+\mathrm{LerchPhi}(e^{-4i\arctan z},1,\frac12)e^{-2i\arctan z}
\\
-\mathrm{LerchPhi}(e^{-4i\arctan z},1,1+\frac{\kappa}{4i})
e^{-4i\arctan z}e^{-\kappa\arctan z}
\Bigr]
\\
=\frac{\sqrt{2}}{1-\nu}
\Bigl[
\Re F_c(\kappa)+\Bigl(\frac{\pi}{2\sqrt{3}}\frac{\nu-1}{\nu+1}
+\frac{2i}{\kappa}\Bigr)+\log iz-2i f^{(\kappa)}(z)
\\
+\mathrm{LerchPhi}(e^{-4i\arctan z},1,1+\frac{\kappa}{4i})
e^{-4i\arctan z}e^{-\kappa\arctan z}
\Bigr].
\end{multline}
Substitution of this expression into \eqref{gen-func-kappa} yields
\begin{multline}
F^{(\kappa)}(z)=V_0^{(\kappa)}\left[\frac{2}{b}\right]^{\frac{1}{2}}
\Bigl[
-\frac{b}{4}-\Bigl(\Re F_c(\kappa)-\frac{b}{4}\Bigr)e^{-\kappa\arctan z}
-\Bigl(\frac{\pi}{2\sqrt{3}}\frac{\nu-1}{\nu+1}+\frac{2i}{\kappa}\Bigr)
-\log iz+2if^{(\kappa)}(z)
\\
-\mathrm{LerchPhi}(e^{-4i\arctan z},1,1+\frac{\kappa}{4i})
e^{-4i\arctan z}e^{-\kappa\arctan z}
\Bigr].
\end{multline}

Now let us consider the generating function
for the $\kappa=0$. Using the fact that
\begin{equation*}\mathrm{LerchPhi}(w,1,1)={}_2F_1(1,1;2;w)=-\frac{\log(1-w)}{w}
\end{equation*}one obtains the following expression
\begin{equation}F^{(0)}(z)=-V_{0}^{(0)}\left[\frac{2}{b}\right]^{\frac{1}{2}}\log(1+z^2).
\label{F:kappa=0}
\end{equation}
\subsection{Normalization of the vectors from continuous spectrum}
The aim of this subsection is to compute the following scalar
product
\begin{equation}(\!(V^{(\kappa)},\,V^{(\kappa')})\!)
=\overline{V}_{0}^{(\kappa)}V_0^{(\kappa)}
+(V^{(\kappa)},\,V^{(\kappa')}).
\end{equation}Substitution of \eqref{cont:V} yields
\begin{multline*}
(\!(V^{(\kappa)},\,V^{(\kappa')})\!)=\overline{V}_{0}^{(\kappa)}V_0^{(\kappa)}
+\overline{A^{(\kappa)}}A^{(\kappa')}(v^{(\kappa)},\,v^{(\kappa')})
-[1-\overline{\nu}(\kappa)]\frac{\overline{V}_0^{(\kappa)}A^{(\kappa')}}{\sqrt{b}}
\Bigl(\mathscr{P}\frac{1}{C'U'-\nu(\kappa)}W,\,v^{(\kappa')}\Bigr)
\\
-[1-\nu(\kappa')]\frac{V_0^{(\kappa')}\overline{A}^{(\kappa)}}{\sqrt{b}}
\Bigl(v^{(\kappa)},\,\mathscr{P}\frac{1}{C'U'-\nu(\kappa')}W\Bigr)
\\
+[1-\overline{\nu}(\kappa)][1-\nu(\kappa')]
\frac{V_0^{(\kappa')}\overline{V}^{(\kappa)}}{b}
\Bigl(\mathscr{P}\frac{1}{C'U'-\nu(\kappa)}W,\,\mathscr{P}\frac{1}{C'U'-\nu(\kappa')}W\Bigr).
\end{multline*}
Substitution of the expressions \eqref{N(kappa)}, \eqref{vWscalar}
and \eqref{A(kappa)} yields
\begin{multline}
=|V_0^{(\kappa)}|^2\Nc(\kappa)\frac{2}{b}\kappa^2\Bigl[
\Re F_c(\kappa)-\frac{b}{4}
\Bigr]^2\delta(\kappa-\kappa')
+\frac{2}{b}
\frac{\overline{V}_0^{(\kappa)}V_0^{(\kappa')}}{\nu(\kappa)-\nu(\kappa')}\Bigl[
\Re F_c(\kappa')(\nu(\kappa)-1)(\nu(\kappa')+1)-(\kappa\leftrightarrow \kappa')
\Bigr]
\\
+[1-\overline{\nu}(\kappa)][1-\nu(\kappa')]\frac{\overline{V}_0^{(\kappa)}V_0^{(\kappa')}}{b}
\Bigl(\mathscr{P}\frac{1}{C'U'-\nu(\kappa)}W,\,\mathscr{P}\frac{1}{C'U'-\nu(\kappa')}W\Bigr).
\label{norm:C}
\end{multline}
Let us now compute the expression
\begin{multline}
\Bigl(\mathscr{P}\frac{1}{C'U'-\nu(\kappa)}W,\,\mathscr{P}\frac{1}{C'U'-\nu(\kappa')}W\Bigr)
\\
=
-2\nu(\kappa)\int_{-\infty}^{\infty}\frac{d\kappa''}{\kappa''}
\frac{[1+\nu(\kappa'')]^2}{\kappa''\Nc(\kappa'')}
\,
\mathscr{P}\frac{1}{\nu(\kappa'')-\nu(\kappa')}
\mathscr{P}\frac{1}{\nu(\kappa'')-\nu(\kappa)}.
\label{norm:cont:A}
\end{multline}
We have to compute this integral very carefully,
because it has a singularity then $\kappa'=\kappa$.
To extract it we will use the following identity
\begin{equation*}\mathscr{P}\frac{1}{x''-x}=\frac{1}{x''-x+i0}+\pi i\delta(x''-x).
\end{equation*}In our case this relation takes a form
\begin{equation*}\mathscr{P}\frac{1}{\nu(\kappa'')-\nu(\kappa)}
=\frac{1}{\nu(\kappa'')-\nu(\kappa)+i0}+\frac{\pi i}{\frac{d\nu}{d\kappa}}\delta(\kappa''-\kappa).
\end{equation*}In principal, one has to understand all these relations as
a specification of the integration contour in \eqref{norm:cont:A}.
Substitution of this relation into \eqref{norm:cont:A} yields
\begin{multline*}
=-2\nu(\kappa)\int_{-\infty}^{\infty}\frac{d\kappa''}{\kappa''}
\frac{[1+\nu(\kappa'')]^2}{\kappa''\Nc(\kappa'')}
\,\left[
\frac{1}{\nu(\kappa'')-\nu(\kappa')+i0}
\frac{1}{\nu(\kappa'')-\nu(\kappa)+i0}\right.
\\
+\frac{\pi i}{\frac{d\nu}{d\kappa}}\delta(\kappa''-\kappa)\frac{1}{\nu(\kappa'')-\nu(\kappa')+i0}
+\frac{\pi i}{\frac{d\nu}{d\kappa'}}\delta(\kappa''-\kappa')\frac{1}{\nu(\kappa'')-\nu(\kappa)+i0}
\\
\left.
-\frac{\pi^2}{\frac{d\nu}{d\kappa}\frac{d\nu}{d\kappa'}}
\delta(\kappa''-\kappa)\delta(\kappa''-\kappa')
\right]
\end{multline*}
Using the relation
\begin{equation*}\frac{2\kappa}{\pi}\Nc(\kappa)\frac{d\nu(\kappa)}{d\kappa}=[1-\nu(\kappa)]^2
\end{equation*}we can simplify this expression
\begin{multline}
=-2\nu(\kappa)\int_{-\infty}^{\infty}\frac{d\kappa''}{\kappa''}
\frac{[1+\nu(\kappa'')]^2}{\kappa''\Nc(\kappa'')}
\,\left[
\frac{1}{\nu(\kappa'')-\nu(\kappa')+i0}
\frac{1}{\nu(\kappa'')-\nu(\kappa)+i0}\right]
\\
+\mathscr{P}\frac{\nu(\kappa)}{\nu(\kappa)-\nu(\kappa')}
\left[\frac{4i}{\kappa'}\frac{1+\nu(\kappa')}{1-\nu(\kappa')}
-\frac{4i}{\kappa}\frac{1+\nu(\kappa)}{1-\nu(\kappa)}
\right]
+\frac{8\Nc(\kappa)}{[1-\nu(\kappa)][1-\overline{\nu}(\kappa)]}
\delta(\kappa-\kappa').
\label{norm:35}
\end{multline}
The integral in the first line can be rewritten as
\begin{multline*}
2\int_{-\infty}^{\infty}\frac{dx''}{x''}
\frac{\sinh\frac{x''}{2}}{\cosh\frac{x''}{2}}
\,\frac{1+2\cosh x''}{[(2\nu'+1)\cosh x''+i\sqrt{3}\sinh x''+2+\nu'+i0]}
\\
\times\frac{1}{
(2\nu+1)\cosh x''+i\sqrt{3}\sinh x''+2+\nu+i0}.
\end{multline*}
This integral has the following poles
\begin{itemize}
\item $x''=\pi i+2\pi i n$, for $n\geqslant 0$;
\item $x''=\frac{\pi \kappa'}{2}+2\pi i n$, for $n\geqslant 1$;
\item $x''=-\frac{\pi \kappa'}{2}+2\pi i n$, for $n\geqslant 1$.
\end{itemize}
Calculation of the residues yields the following expression
for \eqref{norm:35}
\begin{multline}
=2\sum_{n=0}^{\infty}\left[
\frac{1}{n+\frac12}\frac{-2}{(1-\nu')(1-\nu)}
+\mathscr{P}\frac{\nu}{\nu-\nu'}\Bigl(
\frac{1+\nu'}{1-\nu'}\frac{1}{n+1+\frac{\kappa'}{4i}}
-\frac{1+\nu}{1-\nu}\frac{1}{n+1+\frac{\kappa}{4i}}
\Bigr)
\right]
\\
+\mathscr{P}\frac{\nu}{\nu-\nu'}\Bigl(
\frac{1+\nu'}{1-\nu'}\frac{4i}{\kappa'}
-\frac{1+\nu}{1-\nu}\frac{4i}{\kappa}
\Bigr)
+\frac{8\Nc(\kappa)}{[1-\nu(\kappa)][1-\overline{\nu}(\kappa)]}
\delta(\kappa-\kappa').
\\
=\mathscr{P}\frac{2\nu}{\nu-\nu'}\left[
\frac{1+\nu}{1-\nu}\Re F_c(\kappa)
-\frac{1+\nu'}{1-\nu'}\Re F_c(\kappa')
\right]+\frac{8\Nc(\kappa)}{(1-\nu)(1-\overline{\nu})}
\delta(\kappa-\kappa')
.
\end{multline}
Substitution of this expression into \eqref{norm:C} yields
\begin{equation}(\!(V^{(\kappa)},\,V^{(\kappa')})\!)
=\frac{2}{b}|V_0^{(\kappa)}|^2\Nc(\kappa)\left[4+\kappa^2\Bigl(
\Re F_c(\kappa)-\frac{b}{4}
\Bigr)^2\right]\delta(\kappa-\kappa').
\label{NORML:C}
\end{equation}To obtain standard normalization of the continuous spectra one has to put
\begin{equation}V_0^{(\kappa)}=\left[\frac{b}{2}\right]^{\frac12}
\Bigl[\Nc(\kappa)\Bigr]^{-\frac12}\left[4+\kappa^2\Bigl(
\Re F_c(\kappa)-\frac{b}{4}
\Bigr)^2\right]^{-\frac12}.
\label{V0:cont}
\end{equation}

\newpage

\end{document}